\documentclass[reprint]{revtex4-1}
\usepackage{amsfonts}
\usepackage{amsmath}
\usepackage{amssymb}
\usepackage{bm}
\usepackage{color}
\usepackage{graphicx}
\usepackage{hyperref}
\usepackage{mathtools}
\DeclareMathOperator{\arccoth}{arccoth}
\DeclareMathOperator{\diag}{diag}
\DeclareMathOperator{\IFT}{IFT}

\DeclareMathOperator{\Tr}{Tr}

\newcommand{\beq}{\begin{equation}}
\newcommand{\eeq}{\end{equation}}

\begin{document}
\title{Semiclassical analysis of a magnetization plateau in a 2D frustrated ferrimagnet}
\date{\today}
\author{Edward Parker}
\email{tparker@physics.ucsb.edu}
\affiliation{Department of Physics, University of California, Santa Barbara, CA 93106}
\author{Leon Balents}
\affiliation{Kavli Institute for Theoretical Physics, University of California, Santa Barbara, CA 93106}

\begin{abstract}
We use a semiclassical large-$S$ expansion to study a plateau at $1/3$ saturation in the magnetization curve of a frustrated ferrimagnet on a spatially anisotropic kagom\'{e} lattice.  The spins have both ferromagnetic and antiferromagnetic nearest-neighbor Heisenberg couplings, and a frustrating next-nearest-neighbor coupling in one lattice direction.  The magnetization plateau appears at the classical level for a certain range of couplings, and quantum fluctuations significantly broaden it at both ends.  Near the region of the phase diagram where the classical plateau destabilizes, we find an exotic ``chiral liquid'' phase that preserves translational and $U(1)$ spin symmetry, in which bound pairs of magnons with opposite spins are condensed.   We show how this state is obtained naturally from a relativistic field theory formulation.  We comment on the relevance of  the model to the material $\text{Cu}_3\text{V}_2\text{O}_7\text{(OH)}_2 \cdot 2\text{H}_2\text{O}$ (volborthite).
\end{abstract}

\maketitle

\section{Introduction}

Frustrated low-spin magnets provide a rich source of model systems to explore many-body phenomena \cite{BalentsSL,Starykh}.  Perhaps the most manifestly {\em quantum} example observed in many real materials is the magnetization plateau \cite{Rice, Giamarchi}.  This occurs in magnets in applied fields when both the Hamiltonian and the spin state is invariant under rotation about the field axis.  In this situation, the magnetization in the zero temperature limit becomes constant over a finite range of field, forming a plateau, with a quantized rational spin per site.  Through a standard mapping of spins to bosons, a plateau state can be regarded as a ``bosonic Mott insulator,'' and consequently the edges of the plateau can be considered quantum phase transitions, in the category of bosonic Mott transitions.   The precise nature of these transitions depends upon the symmetries of the system, and on the way in which the system is tuned off its plateau, allowing for a complex phase diagram and multiple universality classes for transitions to arise surrounding the plateau state.  This has been explored extensively in the context of the spatially anisotropic triangular lattice \cite{Alicea, ChubukovStarykh, Starykh}, and applied to understand magnetization plateaus in Cs$_2$CuBr$_4$, which realizes that system experimentally \cite{Ono, Tsujii, Fortune}.  

As they typically arise (at least in magnets with weak anisotropy) from quantum fluctuation effects,  magnetization plateaus in such systems are usually narrow -- as is indeed the case in Cs$_2$CuBr$_4$.  However, one can contemplate situations in which a {\em wide} plateau occurs.  This happens trivially in some highly anisotropic, i.e. Ising-like, magnets, because the spins behave discretely and so only adjust when a barrier to flipping is overcome.  However, in Heisenberg-like magnets, a wide plateau is more interesting.  Here we consider one mechanism for this case, based on the proximity to a {\em ferrimagnetic} state.  In a ferrimagnet, spins spontaneously order at zero field in a collinear fashion with a nonzero net moment but spins oriented both parallel and antiparallel to the polarization axis: thus a ferrimagnet has a uniform magnetization which is less than the maximum saturated ferromagnetic one.  Application of an infinitesimal external field fixes the polarization axis, leading to a plateau of magnetization which extends down to zero field. 

In this paper, we consider how such a plateau is modified by frustration which removes the zero field ferrimagnetic state but leaves the system nearby in phase space.  We build a prototypical model of such a ``frustrated ferrimagnet'' by starting  with a well-studied but very rich one-dimensional model, the frustrated ferromagnetic spin $S=1/2$ chain, and coupling this antiferromagnetically to a set of interstitial spins between the chains, which converts the (nearby) ferromagnetism of the chains to (nearby) ferrimagnetism.  The result is a rich phase diagram, dependent upon exchange interactions and field, within which we particularly explore the plateau, its stability, and the surrounding quantum phase transitions.

The model we study has the geometry of an anisotropic kagom\'e lattice, and is inspired by experiments \cite{Ishikawa} on $\text{Cu}_3\text{V}_2\text{O}_7\text{(OH)}_2 \cdot 2\text{H}_2\text{O}$ (volborthite), which shows an extremely broad magnetization plateau at $1/3$ saturation between applied fields of 28 T to over 120 T \cite{Yamashita}.  Early density functional calculations \cite{Janson10} suggested that volborthite could be described by the model we study, though a more recent investigation \cite{Janson16} using similar methods obtains a different set of exchange interactions with less symmetry that no longer has a coupled chain structure.  We give our own opinion of the relevance of the coupled chain model to volborthite in Sec.~\ref{Experiment}.  Narrower $1/3$ magnetization plateaus have also been proposed in isotropic nearest-neighbor kagom\'{e} models with Heisenberg \cite{Schulenburg, Nishimoto, Capponi} and $XXZ$ \cite{Damle, Zhu} couplings.  In the majority of this paper, we study the model for its own sake, and in particular explore its physics in the semiclassical limit, and how this connects to universal field theoretic descriptions.  

Our main results are as follows.  We indeed find a broad magnetization plateau, and determine its boundary rather generally.  We obtain theories of the quantum phase transitions off the plateau, which are reminiscent of those for the well-studied Mott transition of bosons in a periodic potential \cite{Fisher}.  The transitions from the plateau driven by increasing and decreasing field are controlled by ``nonrelativistic'' field theories with dynamical critical exponent $z=2$, and physically describe the condensation of magnons with a definite spin quantum number ($S^z=+1$ {\em or} $S^z=-1$ for increasing or decreasing field, respectively).  Near the termination of the plateau with increasing frustrating exchange, we find a regime in which {\em both} $S^z=+1$ {\em and} $S^z=-1$ magnons approach low energy, leading to a relativistic $z=1$ field theory. These theories differ from those in the simple boson Mott case by additional ``flavors'' attached to the bosons, due to the sublattice and momentum/valley degrees of freedom of our frustrated model.  These differences lead to some different physical consequences, particularly in the relativistic regime.  There, we find an ``excitonic'' instability pre-empting the na\"ive relativistic Bose condensation, which results in the establishment of composite {\em chiral} order at the end of the plateau.  This result is quite similar to one obtained for the (simpler) anisotropic triangular lattice in Ref.~\onlinecite{ChubukovStarykh}.  Our methodology is however somewhat different, and shows how both our and this earlier result can be understood in standard field theoretic terms from the relativistic boson formulation.

\begin{figure}
\includegraphics[page=1,width=\columnwidth]{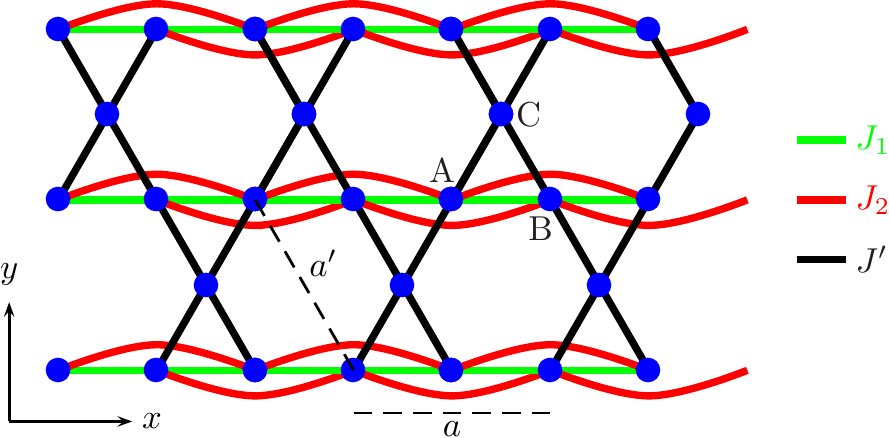}
\caption{Proposed Hamiltonian for volborthite.  The blue dots represent spin-1/2 copper ions and the line segments represent Heisenberg couplings.  $J_1 < 0$ is ferromagnetic while $J_2 > 0$ and $J' > 0$ are antiferromagnetic.  The distances between adjacent unit cells is slightly anisotropic, with $a = 5.84$ \AA\ and $a' = 6.07$ \AA\ \cite{Ishikawa}.  Capital letters label the sublattices.} \label{Hamiltonian diagram}
\end{figure}

\emph{The model}.  We consider the Hamiltonian
\begin{align}
H =& \sum_{\langle i j \rangle \parallel \hat{x}} J_1\, \bm{S}_i \cdot \bm{S}_j + \sum_{\langle \langle i j \rangle \rangle \parallel \hat{x}} J_2\, \bm{S}_i \cdot \bm{S}_j \label{Hamiltonian} \\
&+ \sum_{\langle i j \rangle \nparallel \hat{x}} J'\, \bm{S}_i \cdot \bm{S}_j - \sum_i h\, S_i^z \nonumber
\end{align}
illustrated in Fig.~\ref{Hamiltonian diagram}, where $\bm{S}_i$ is a spin-$S$ operator at site $i$, $\langle i j \rangle$ denotes nearest-neighbor sites and $\langle \langle i j \rangle \rangle$ next-nearest neighbors, and $h$ is an applied field in the $z$-direction in spin space.   In the limit $J_2 \ll J_1, J'$, the couplings are unfrustrated, and the resulting $1/3$ magnetization plateau can be interpreted semiclassically with the chain spins aligned with the field and the interstitial spins aligned against it (with quantum fluctuations from the antiferromagnetic couplings reducing the moments' magnitude from full polarization).  This ferrimagnetic state is sometimes called the up-up-down or UUD state \cite{Alicea}.

In the limit $J' \ll J_1, J_2$, the chains decouple and the system reduces to the frustrated ferromagnetic $S=1/2$ $J_1$-$J_2$ chain, which describes systems such as $\text{LiCuVO}_4$ and has been the subject of much recent study \cite{Chubukov, Heidrich-Meisner, Kuzian, Kecke, Hikihara, Sudan, Sato11, Sato13, Balents}.  This model has a very complex phase diagram with at least six different ordering behaviors \cite{Hikihara, Sudan}.  However, we focus here on the semiclassical description, which we believe is rather accurate when $J'$ is not extremely small.

The paper is organized as follows.  Sec.~\ref{Classical} explores the classical $T = 0$ phase diagram.  Sec.~\ref{Spin waves} reformulates the problem in terms of canonical bosonic magnons.  Sec.~\ref{Corrections} examines the $o(S^0)$ contributions to the magnon self-energies and the resulting shift in the plateau's classical critical fields; we find that the quantum corrections broaden the plateau in both directions.  Sec.~\ref{Interactions} considers the effect of magnon interactions.  Sec.~\ref{RG} studies the renormalization-group (RG) flow of the couplings in the exotic ``chiral liquid'' phase mentioned above.  Sec.~\ref{Wavefunction} examines the microscopic wavefunction in this phase; we find it is characterized by a ``spin current'' induced by exciton condensation.  Sec.~\ref{Triangular lattice} compares our results to previously studied models, and Sec.~\ref{Experiment} to the volborthite experimental data.  Sec.~\ref{Conclusion} concludes.

\section{Classical $T = 0$ phase diagram \label{Classical}}

\begin{figure}
\includegraphics[page=2]{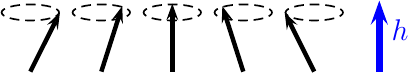}
\caption{Classical cone state precessing in real space about an applied field $\bm{h}$.} \label{Cone}
\end{figure}

For a classical Heisenberg magnet whose couplings have the translational symmetry of a Bravais lattice, the Luttinger-Tisza theorem \cite{Luttinger, Kaplan} gives that the ground state takes the form of a ``cone'' state (sometimes called ``helical'' or ``umbrella'' state) illustrated in Fig.~\ref{Cone}, in which the spins spatially precess about the applied field with a uniform (commensurate or incommensurate) wave vector $\bm{k}$:
\beq \label{LT}
\bm{S_x} = (\sin \theta \cos(\bm{k} \cdot \bm{x} - \delta), \sin \theta \sin(\bm{k} \cdot \bm{x} - \delta), \cos \theta),
\eeq
where $\theta$ is the spin cone's opening angle from the field.  The global $U(1)$ rotational symmetry about the applied field shifts the value of $\delta$.

\begin{figure}
\includegraphics[trim={0 2.9cm 0 2.8cm}, clip, width=\columnwidth]{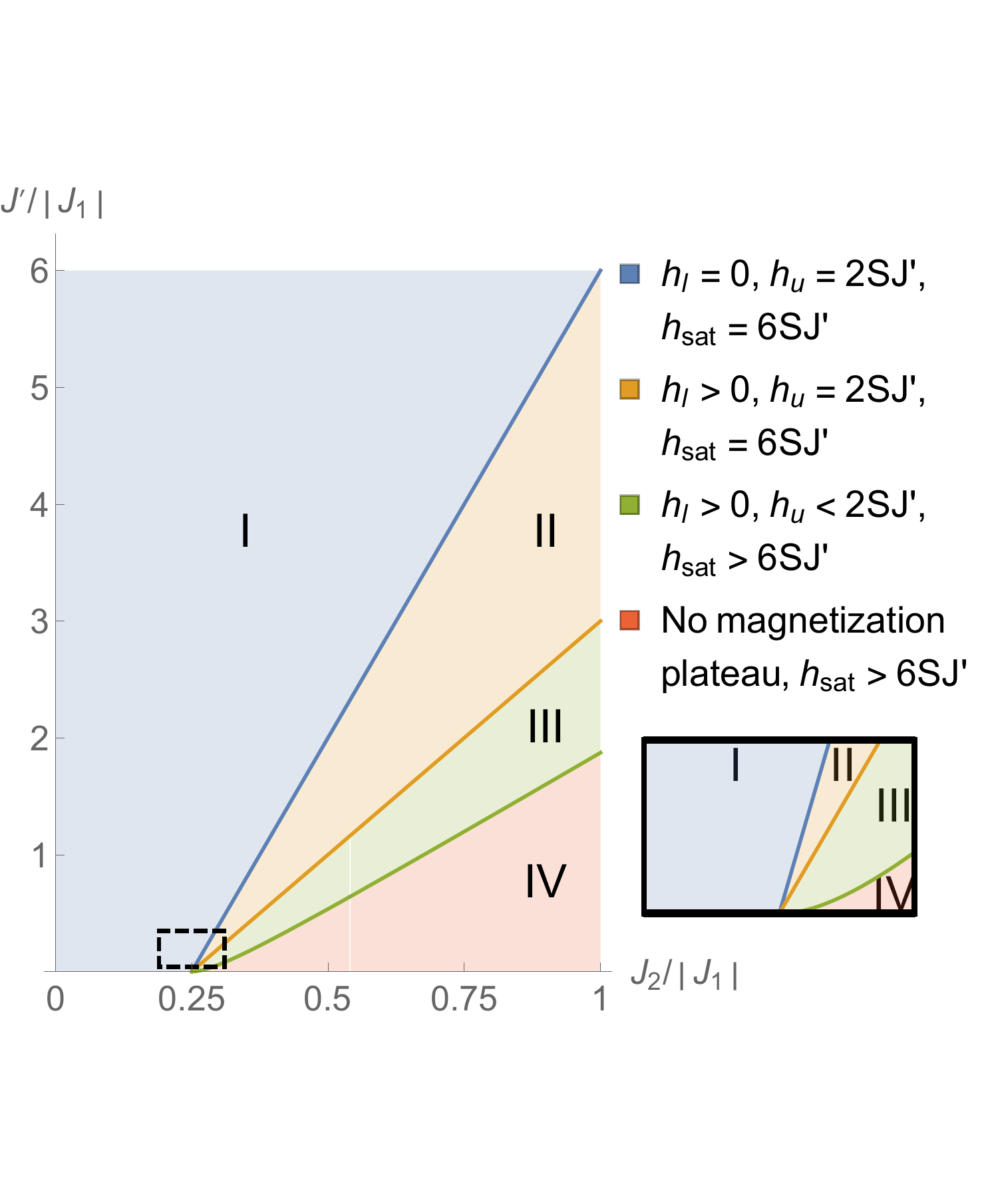}
\caption{Projection of classical $T=0$ phase diagram onto the $h = 0$ plane.  $h_l$ and $h_u$ are the plateau's lower and upper critical fields, respectively, and $h_\text{sat}$ is the saturation field.  The lower and upper boundaries of region II are given by $J' = J_1 + 4 J_2$ and $J' = 2(J_1 + 4 J_2)$ respectively.  The ``stabilization curve'' boundary between regions III and IV is linear for large $J'$ and quadratic near the Lifshitz point $(J_2 = |J_1| / 4, J' = 0)$ (shown inset) \cite{app}.} \label{LT phase diagram}
\end{figure}

The Luttinger-Tisza theorem can be generalized to certain lattices with multiple spins per primitive unit cell \cite{Litvin, Xiong}, but our model has three inequivalent spins per primitive unit cell and the theorem does not apply.  We therefore make an ansatz generalizing the form \eqref{LT} so that the cone opening angle $\theta$ depends on $y$, reflecting the fact that the antiferromagnetic $J'$ coupling tends to make the cones on the chains and on the interstitial spins open in opposite directions.  From this assumption, we can prove that the energy-minimizing spin configurations on each chain must be identical, and the spin configurations on each row of interstiatial spins must also be identical (App.~\ref{Classical appendix}).  Therefore, all the chains have the same opening angle $\theta_c$, all the interstitial spins have the same opening angle $\theta_{IS}$, and $\bm{k}$ points in the $x$-direction.  For a system of $N$ spins, \eqref{Hamiltonian} then reduces to

\begin{align} \label{LTGSE}
\frac{E}{N} =& \frac{2}{3} S^2 \left[ \vphantom{\frac{1}{2}} J_1 \left( \sin^2 \theta_c \cos \Delta \phi + \cos^2 \theta_c \right) \right. \\
& \hspace{25pt} + J_2 \left( \sin^2 \theta_c \cos (2 \Delta \phi) + \cos^2 \theta_c \right) \nonumber \\
& \hspace{25pt} + 2 J' \left( \sin \theta_c \sin \theta_{IS} \cos(\Delta \phi / 2) + \cos \theta_c \cos \theta_{IS} \right) \nonumber \\
& \hspace{25pt} \left. -\frac{h}{S} \left(\cos \theta_c + \frac{1}{2} \cos \theta_{IS} \right) \right], \nonumber
\end{align}
where $\Delta \phi := k_x a / 2$ is the azimuthal angle between two adjacent chain spins.  The sign of $\Delta \phi$ gives the orientation of the cone's precession and spontaneously breaks the chiral symmetry of the Hamiltonian.  By extremizing \eqref{LTGSE} with respect to $\theta_c$, $\theta_{IS}$, and $\Delta \phi$, we can find the ground-state spin configuration (within our ansatz) for arbitrary Hamiltonian parameters \cite{app}.

We are primarily interested in couplings which produce a 1/3 magnetization plateau -- i.e. for a finite range $h \in [h_l, h_u]$, the ground state has all the chain spins aligned with the field $(\theta_c = 0)$ and all the interstitial spins antialigned $(\theta_{IS} = \pi)$.  The energy \eqref{LTGSE} is always stationary at this point; to find the plateau phase, we must find the Hamiltonian parameters for which $(\theta_c = 0, \theta_{IS} = \pi)$ is a global minimum \cite{app}.  (The plateau is also stable against position-dependent perturbations for these parameters; see Sec.~\ref{Spin waves}.)

\begin{figure}
\includegraphics[width=\columnwidth]{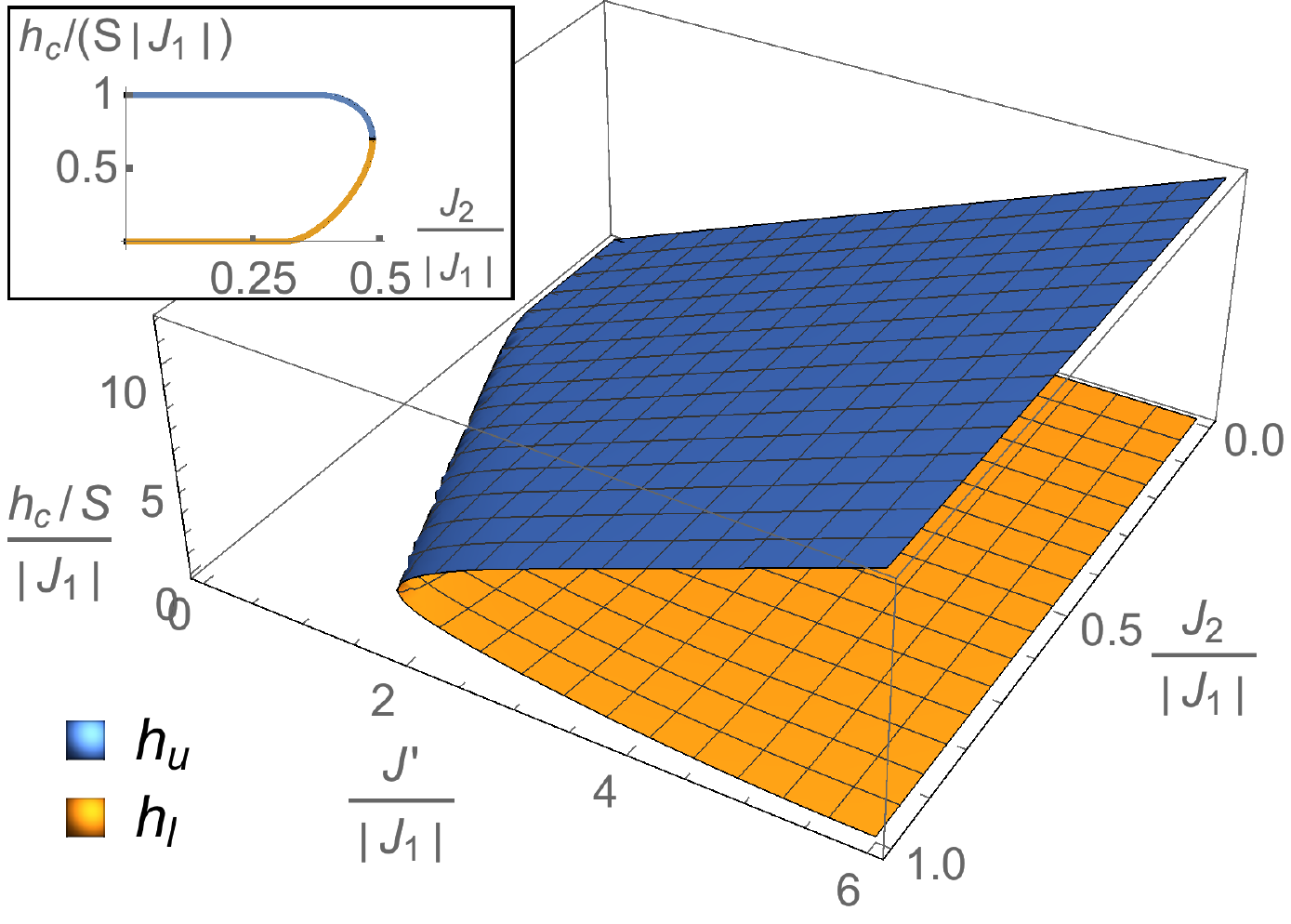}
\caption{Classical $T = 0$ phase diagram including $h$.  The magnetization plateau lies between the two sheets.  The saturation field (not shown) is quite close to planar, with a steeper slope than $h_u$.  The two sheets meet along the stabilization curve.  Inset: a cross-section at $J' = 0.5 |J_1|$.} \label{3D LT phase diagram}
\end{figure}

The results of this stability analysis are illustrated in Fig.~\ref{LT phase diagram}.  In region I, the plateau appears at infinitesimal field and is stable up to $h_u = 2 S J'$.  In region II, $h_l$ comes ``unpinned'' and becomes strictly positive, varying continuously with all three couplings.  In region III, $h_u$ comes unpinned from $2S J'$ and $h_\text{sat}$ from $6S J'$ as well, and both vary continuously with all three couplings.  $h_l$ and $h_u$ meet at the boundary with region IV, which does not support a plateau.  We will refer to this boundary between regions III and IV as the ``stabilization curve.''  For $J_2 \ll J'$, the plateau is so slightly frustrated that it appears at infinitesimal field, while for $J_2 \gg J'$, the plateau is so strongly frustrated that it is not stable at any field.  All four regions meet at the Lifshitz point $\left( J_2 = -\frac{1}{4} J_1, J' = 0 \right)$ -- a highly degenerate critical point near which the quantum phase diagram is extremely complex and unusual even along $J' = 0$ \cite{Balents}.  The expressions for the critical fields and phase boundaries are very complicated but can all be found in closed form \cite{app}.  For couplings away from the stabilization curve, the off-plateau portions of the magnetization curve are close to (but not exactly) linear, as shown in Fig.~\ref{MagCurve}; for couplings at the stabilization curve, the magnetization curve's 1/3 plateau degenerates to a saddle point. 

In Fig.~\ref{3D LT phase diagram}, we introduce the applied field on a new axis and show the three-dimensional phase diagram.  We see that for any fixed $J'$, the phase diagram has a ``lobe'' shape and the upper and lower critical fields do not depend on $J_2$ for small $J_2$.

\section{Spin-wave formulation \label{Spin waves}}

To study quantum corrections to the classical magnetization plateau, we can reformulate \eqref{Hamiltonian} in terms of spin-1 bosonic magnons, where the magnon vacuum is the classical UUD state.  We define canonical bosonic ladder operators $a_i$, $b_i$, and $c_i$ which annihilate magnons on the $A$, $B$, and $C$ sublattices respectively of the primitive unit cell $i$.  We use the standard Holstein-Primakoff transformation \cite{HP} $S_A^+ = \sqrt{2S - n^a}\, a,\ S_A^z = S - n^a$ and similarly for the $B$ sublattice, but $S_C^+ = c^\dag \sqrt{2S - n^c},\ S_c^z = -S + n^c$, where $n^a := a^\dag a$ and similarly for $n^b$ and $n^c$.  The $a$ and $b$ magnons have spin $S^z = -1$ and the $c$ magnon has $S^z = +1$.

In the semiclassical regime where $S$ is large, we can expand \eqref{Hamiltonian} in powers of $S$:
\beq \label{HSeries}
\frac{H}{N/3} = S^2 E_0 + S H^{(2)} + H^{(4)} + o(1/S),
\eeq
where $N$ is the number of spins and $H^{(n)}$ denotes the terms in the Hamiltonian with $n$ magnon ladder operators.  $E_0 = 2(J_1 + J_2 - 2 J') - h/S$ is the classical energy of the plateau.  $H^{(2)}$ gives the free magnon contribution, or equivalently, the energies of the classical perturbations about the plateau.  $H^{(4)}$ gives quantum corrections due to magnon interactions.

We assemble the momentum-basis ladder operators (see App.~\ref{Spin-wave appendix} for Fourier conventions) into a column vector
\beq
\Psi(k) := \left( \begin{array}{c} a(k) \\ b(k) \\ c^\dag(-k) \end{array} \right).
\eeq
Then up to an unimportant shift in the zero-point energy, $H^{(2)} = \sum_k \Psi^\dag(k)\, M(k)\, \Psi(k)$, where $M(k)$ is a matrix given in \eqref{Mk} that depends implicitly on the Hamiltonian parameters.

\emph{Bogoliubov transformation}.  To diagonalize $M(k)$, we must find a $k$-dependent $3 \times 3$ Bogoliubov transformation $T(k)$ of the eigenstate ladder operators $\tilde{\Psi}(k) = \left( \tilde{a}(k)\ \ \tilde{b}(k)\ \ \tilde{c}^\dag(-k) \right)^T$ such that $\Psi(k) = T(k) \tilde{\Psi}(k)$.  Under this transformation, $\Psi^\dag M \Psi = \tilde{\Psi}^\dag T^\dag M T \tilde{\Psi}$, so we want $D(k) := T^\dag(k) M(k) T(k)$ to be diagonal.

The canonical commutation relations can be expressed in terms of a matrix of commutators $g_{\alpha \beta} := \left[ \Psi_\alpha(k), \Psi_\beta^\dag(k) \right] \rightarrow \diag(1, 1, -1)$.  The Bogoliubov transformation must preserve these commutation relations, so we can think of $g$ as a ``metric'' which $T(k)$ must preserve in the same way that Lorentz transformations preserve the Minkowski metric:
\beq
\left[ T_{\gamma \alpha} \tilde{\Psi}_\alpha, \tilde{\Psi}_\beta^\dag T^\dag_{\beta \delta} \right] = T_{\gamma \alpha} g_{\alpha \beta} T_{\beta \delta}^\dag = g_{\gamma \delta}
\eeq
or $TgT^\dag = g$.  Thus $T(k)$ must belong to the fundamental representation of $U(2, 1)$.

The spin structure of the eigenstate ladder operators is given by
\begin{align}
S^z_\text{tot} &= \frac{1}{3} N S - \sum_k \left( n_k^a + n_k^b - n_k^c \right) \nonumber \\
&= \frac{1}{3} N (S - 1) - \sum_k \Psi^\dag(k) g \Psi(k) \\
&= \frac{1}{3} N (S - 1) - \sum_k \tilde{\Psi}^\dag(k) g \tilde{\Psi}(k) \nonumber \\
&= \frac{1}{3} N S - \sum_k \left( n_k^{\tilde{a}} + n_k^{\tilde{b}} - n_k^{\tilde{c}} \right). \nonumber
\end{align}
The Bogoliubov transformation preserves the fact that the $\tilde{a}$ and $\tilde{b}$ magnons carry spin $S^z = -1$ and the $\tilde{c}$ magnon carries spin $S^z = +1$, so the applied field acts as a chemical potential for magnon eigenstates.  When the plateau is stable, there is a gap to create a magnon (flip a spin).  At applied field $h_u$, the gap to create a $\tilde{c}$ magnon closes, $\tilde{c}$ magnon condensation spontaneously breaks the $U(1)$ symmetry preserving $S^z_\text{tot}$, and the plateau destabilizes to a magnetization higher than $\frac{1}{3} m_\text{sat}$.  At applied field $h_l$, the $\tilde{a}$ or $\tilde{b}$ magnons condense instead, and the plateau destabilizes to a lower magnetization.

From this perspective, the phase diagram in Fig.~\ref{3D LT phase diagram} can be throught of as describing a bosonic system with three phases: a gapped, fully saturated phase at high field (not shown) and a gapped plateau phase between the critical fields, separated by a gapless superfluid-like phase.  The lobe-shaped dependence of the gapped plateau phase on the chemical potential (depicted in the inset to Fig.~\ref{3D LT phase diagram}) closely resembles the commensurate-density Mott-insulator phase adjacent to a gapless superfluid phase in a system of repulsively interacting bosons in an ordered potential \cite{Fisher}.  In our model, the role of the boson hopping term is played by the $J_2$ coupling that frustrates the plateau.

To calculate $T(k)$, we use
\beq \label{calcT}
T^\dag M T = D \implies T g T^\dag M T = g M T = T g D.
\eeq
The diagonal elements $\{ \epsilon_{\tilde{a}}(k), \epsilon_{\tilde{b}}(k), \epsilon_{\tilde{c}}(-k) \} / S$ of $D(k)$ give the free magnon energies.  So $\hat{D}(k) := g D(k)$ is also diagonal and has elements $\{ \epsilon_{\tilde{a}}(k), \epsilon_{\tilde{b}}(k), -\epsilon_{\tilde{c}}(-k) \} / S$, and $(g M) T = T \hat{D}$ implies that the columns of $T(k)$ are the eigenvectors of $g M(k)$.  The magnons must be gapped for the classical magnetization plateau to be stable, so all three $\epsilon(k)$'s must be strictly postive.

\emph{Free magnons with $k_y = 0$}.  For general $k$, the eigenvalues of $gM(k)$ are very complicated.  But by symmetry, the classical instabilities must occur along the $k_x$-axis, where the eigenvalues become much simpler.  One of the $S^z = -1$ eigenmodes, which we will choose to correspond to $\tilde{a}$, is always gapped for any couplings that permit a plateau, so it does not affect the critical fields \cite{app}.  In terms of the position-basis ladder operators,
\beq \label{atilde}
\tilde{a}(k_x, 0) = \frac{1}{\sqrt{2}} \left( a(k_x, 0) - e^{- \frac{1}{2} i\, k_x}\, b(k_x, 0) \right),
\eeq
so the gapped mode is simply a single plane wave traveling down the $J_1$-$J_2$ chains.  The $\tilde{b}$ and $\tilde{c}$ operators have energies 
\beq \label{bcEnergies}
\epsilon_{\tilde{b}/\tilde{c}}(k_x, 0) = S \sqrt{f_1(k_x)} \pm (h - S f_2(k_x)),
\eeq
where $+$ corresponds to $\tilde{b}$, $-$ to $\tilde{c}$, and $f_1(k_x)$ and $f_2(k_x)$ are defined in \eqref{f1f2}.  The classical plateau is only stable if $f_1(k_x)$ is strictly positive, so the stabilization curve is given by the parameters at which $f_1(k_x)$ crosses zero.  Since both energies must be positive for the plateau to be stable, the critical fields are given by $h_l = S \max_{k_x} \left( f_2(k_x) - \sqrt{f_1(k_x)} \right)$ and $h_u = S \min_{k_x} \left( f_2(k_x) + \sqrt{f_1(k_x)} \right)$.  The resulting critical fields and instability wave vectors $k_l$ and $k_u$ for the lower and upper critical fields \cite{app} agree exactly with those found via our generalized Luttinger-Tisza ansatz, so the plateau phase shown in Fig.~\ref{3D LT phase diagram} is stable against arbitrary position-dependent perturbations, not just perturbations that respect the ansatz.

\emph{Bogoliubov transformation for $k_y = 0$}.  The Bogoliubov transformation along the $k_x$-axis takes a particularly simple form
\beq \label{T}
T(k_x, 0) =  \frac{1}{\sqrt{2}} \left( \begin{array}{ccc}
1 & e^{-\frac{1}{4} i k_x} \cosh \xi & e^{-\frac{1}{4} i k_x} \sinh \xi \\
-e^{\frac{1}{2} i\, k_x} & e^{\frac{1}{4} i\, k_x} \cosh \xi & e^{ \frac{1}{4} i\, k_x} \sinh \xi \\
0 & \sqrt{2}\, \sinh \xi & \sqrt{2}\, \cosh \xi
\end{array} \right)
\eeq
\cite{app} which depends on the couplings only through the dimensionless parameter $\xi(k_x)$ defined by
\beq \label{xi}
\tanh(2 \xi(k_x)) = \frac{2 \sqrt{2} J' \cos \left(\frac{k_x}{4} \right)}{J_1 \left(1 - \cos \left( \frac{k_x}{2} \right) \right) + J_2 (1 - \cos k_x) - 3 J'}.
\eeq

\emph{Spin waves at the saturation field}.  We can perform a similar expansion of the Hamiltonian about the fully polarized state.  We define $\Psi_\text{sat}(k) := \left( \begin{array}{ccc} a(k) & b(k) & c(k) \end{array} \right)^T$ and find that the quadratic Hamiltonian $H^{(2)} = \sum_k \Psi_\text{sat}^\dag M_\text{sat} \Psi_\text{sat}$, where the matrix $M_\text{sat}(k)$ is given in App.~\ref{Spin-wave appendix}.  This ferromagnetic case is much more straightforward than the ferrimagnetic case considered above, because the Hamiltonian preserves total magnon number and the Bogoliubov transformation is unitary rather than being an element of $U(2,1)$.

In the case $k_y = 0$, one magnon mode is given by \eqref{atilde} and again lives only on the $J_1$-$J_2$ chains.  The energy bands $\epsilon_i(k_x, 0)$ are given in App.~\ref{Spin-wave appendix}.  The Bogoliubov transformation is
\beq
T(k_x, 0) =  \frac{1}{\sqrt{2}} \left( \begin{array}{ccc}
1 & e^{-\frac{1}{4} i k_x} \cos \theta & e^{-\frac{1}{4} i k_x}\sin \theta \\
-e^{\frac{1}{2} i\, k_x} & e^{ \frac{1}{4} i\, k_x} \cos \theta & e^{\frac{1}{4} i\, k_x} \sin \theta \\
0 & \sqrt{2}\, \sin \theta & -\sqrt{2}\, \cos \theta
\end{array} \right)
\eeq
\cite{app} where
\beq
\tan(2 \theta(k_x)) = \frac{2 \sqrt{2} J' \cos \left( \frac{k_x}{4} \right)}{-J_1 \left(1 - \cos \left(\frac{k_x}{2} \right) \right) - J_2 (1 - \cos k_x) + J'}.
\eeq

\section{Magnon self-energy corrections to plateau critical fields \label{Corrections}}

So far our analysis has been essentially classical.  To incorporate the effects of quantum fluctuations, we must calculate the magnon self-energies $\Delta \epsilon_i$ due to interactions, which will shift the critical fields at which the magnons condense and destabilize the plateau.  If $S$ is large, then we can treat the interaction term $H^{(4)}$ in \eqref{HSeries} perturbatively and neglect higher interactions.  The leading-order contributions in $1/S$ to the self-energies are simply the expectation values of $H^{(4)}$ (which are $o(S^0)$) with respect to the free-magnon eigenstates.  To find the corrections to the critical fields, we only need the shifts in the ground-state energies, so we only consider the free magnons $\tilde{b}(k_l)$ and $\tilde{c}(k_u)$ at their noninteracting ground-state momenta (the shifts in $k_l$ and $k_u$ themselves are second-order effects that we neglect).

\begin{figure*}
\includegraphics[width=\textwidth]{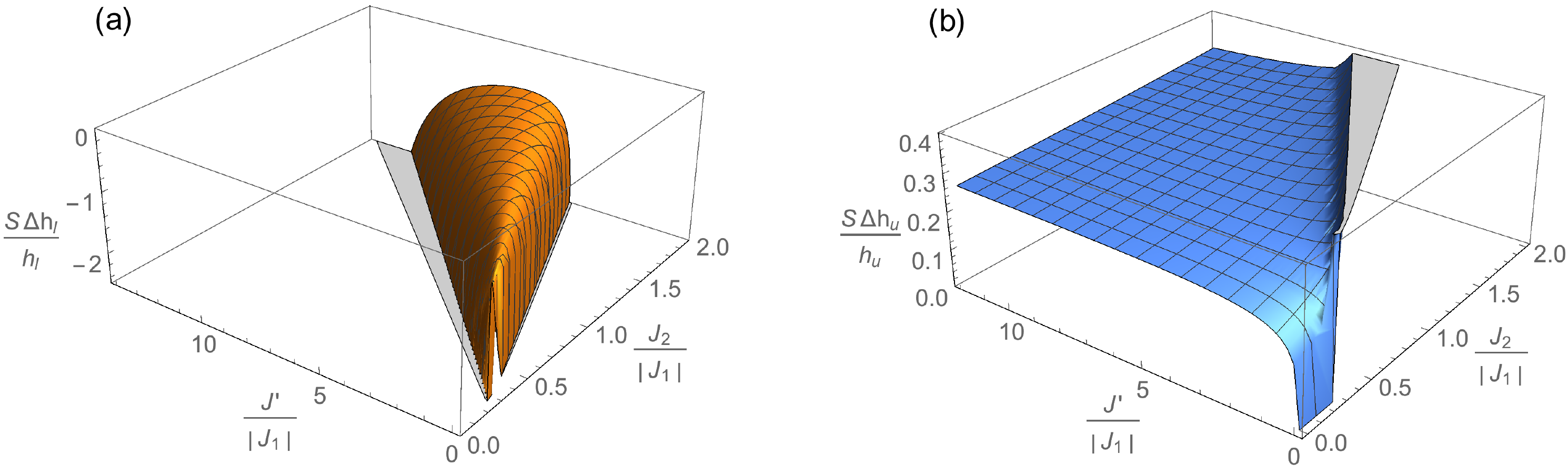}
\caption{Relative leading-order corrections from quantum fluctuations to the plateau's (a) lower critical field and (b) upper critical field.  The divergence of $\Delta h_l / h_l$ at $J' = 2(J_1 + 4 J_2)$ at the left-hand side of (a) reflects that the magnons' self-energy goes to zero more slowly than their noninteracting energy.  Both critical fields also diverge at the stabilization curve because the magnons begin strongly interacting and the single-magnon picture breaks down (discussed in Sec.~\ref{Interactions}).} \label{CorrectionsPlot}
\end{figure*}

One way to evaluate these expectation values is the technique known as Oguchi's corrections \cite{Oguchi}: we first Fourier transform $H^{(4)}$, then perform the Bogoliubov transformation on all the ladder operators (yielding 3420 different quartic interaction terms!), and finally normal-order $H^{(4)}$ \emph{after} the Bogoliubov transformation.  The normal-ordered $H^{(4)}$ has zero expectation value with respect to any one-particle state, but the normal-ordering process will result in an $o(S^0)$ quadratic Hamiltonian which can be expressed in the form $\Delta H^{(2)} = \sum_k \tilde{\Psi}^\dag(k)\, \Delta M(k)\, \tilde{\Psi}(k)$ plus an unimportant zero-point-energy shift.  The diagonal elements of the matrix $M(k)$ give the leading-order self-energies.  (See App.~\ref{Corrections app} for details of the calculation; expressions for the magnon self-energies are given in \eqref{delta es}.)  The leading-order quantum corrections to the plateau critical fields are given by $\Delta h_l = -\Delta \epsilon_{\tilde{b}}(k_l)$ and $\Delta h_u = \Delta \epsilon_{\tilde{c}}(k_u)$.

\begin{figure}
\includegraphics[trim={1.19cm 0 0.92cm 0}, clip, width=\columnwidth]{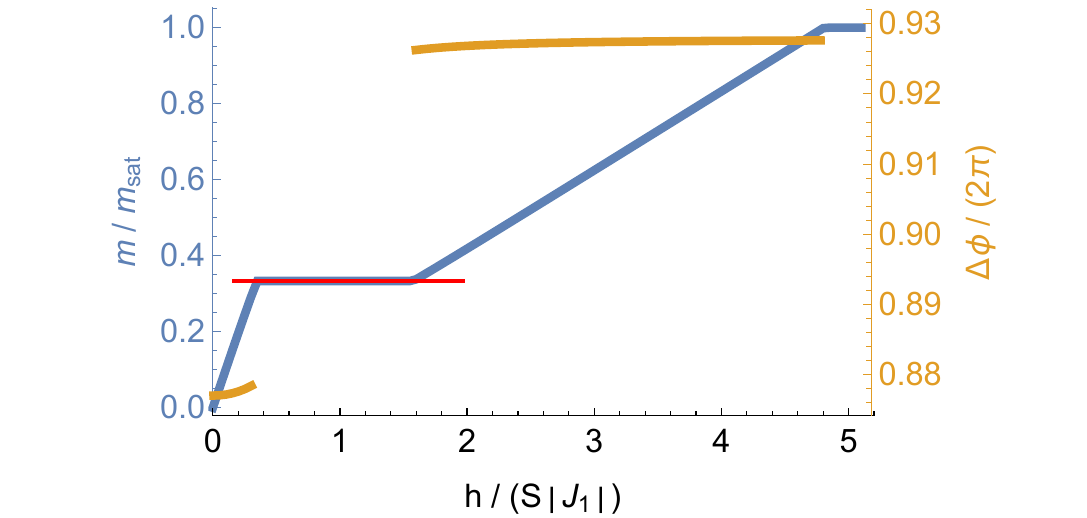}
\caption{The blue curve (corresponding to the left-hand vertical axis) is the classical magnetization curve given by \eqref{mofh} for $J_2 = 0.5 |J_1|$, $J' = 0.8 |J_1|$.  The thin red line indicates the broader plateau incorporating the leading quantum corrections for $S = 3/2$.  Off the plateaus, the gold curve (corresponding to the right-hand vertical axis) shows the cone ground state's classical precession wave number $\Delta \phi$ for the same couplings.}
\label{MagCurve}
\end{figure}

The relative corrections $S\, \Delta h_c / h_c$ to the classical critical fields are plotted in Fig.~\ref{CorrectionsPlot}.  We see that quantum fluctuations broaden the plateau at both the low and high fields.  This result accords with the general tendency of quantum fluctuations to favor collinear states via order-by-disorder effects \cite{Henley}.  The perturbative expansion in $1/S$ is valid if $|\Delta h_c| \ll h_c$, or equivalently if the plotted quantity $S |\Delta h_c| / h_c \ll S$.  Fig.~\ref{MagCurve} shows a representative classical magnetization curve and the quantum corrections' broadening of the plateau.

For $J' \geq 2(J_1 + 4 J_2)$, both $h_l$ and $\Delta h_l$ are zero.  This result is nonperturbative, because at $h = 0$ the plateau spontaneously breaks the Hamiltonian's $SU(2)$ spin symmetry, so Goldstone's theorem prevents the magnon interactions from opening a gap.  As $J' \rightarrow 2(J_1 + 4 J_2)^-$, the classical lower critical field vanishes quadratically \cite{app}, but the leading-order quantum corrections only vanish linearly, so their ratio diverges as $1/(2(J_1 + 4 J_2) - J')$ and the magnons begin interacting strongly, so our perturbative treatment breaks down.  For $J_2 \gg |J_1|$, the relative correction $S \Delta h_l / h_l$ is fairly flat at about $-0.73$ in between the stabilization curve and $2(J_1 + 4 J_2)$.  The divergence at the stabilization curve is discussed below.

The quantum corrections to the upper critical field $h_u$ are much more stable.  For $J' \geq J_1 + 4 J_2$, the relative correction $S \Delta h_u / h_u$ is approximately $0.289$.  The correction increases with both $J_2$ and $J'$, but the dependence is very weak (for example, at $J' = 12 |J_1|$, the relative corrections only increases from $0.269$ at $J_2 = 0$ to $0.306$ at $J_2 = 2 |J_1|$).  For $J' < J_1 + 4 J_2$ the corrections depend more strongly on $J_2$ and $J'$, but remain fairly small away from the stabilization curve.

The self-energies of both the $\tilde{b}$ and $\tilde{c}$ magnons diverge to $+\infty$ at the stabilization curve, because both parameters $\xi(k_l)$ and $\xi(k_u)$ diverge to $-\infty$.  From \eqref{T}, this means that the Bogoliubov coherence factors for the condensed magnons diverge (as $(J' - J'_c)^{-1/4}$) while those for the gapped magnon do not \cite{app}.  As we discuss in Section~\ref{Interactions}, this divergence occurs because the magnon interactions become superrenormalizable near the stabilization curve.

\section{Magnon interactions \label{Interactions}}

\emph{Away from the stabilization curve.}  Strongly gapped modes do not contribute to magnon condensation for weak interactions, so near the critical fields we need only consider interactions between near-gapless modes.  Away from the stabilization curve, only one species of magnon condenses: $\tilde{b}$ at the lower and $\tilde{c}$ at the upper critical field.  The condensed magnons have momentum $(\pm k_c, 0)$ (where $k_c \geq 0$ solves \eqref{k0}) and can be considered ``right movers'' and ``left movers.''  We consider ``parallel'' interactions between either two right movers or two left movers, and ``antiparallel'' (center-of-mass) interactions between a right mover and a left mover.

At the lower critical field, we consider the quartic interaction terms containing $\tilde{b}^\dag(\pm k_l) \tilde{b}^\dag(\pm k_l) \tilde{b}(\pm k_l) \tilde{b}(\pm k_l)$, where the $\pm$'s are independent except for overall momentum conservation.  We calculate interaction energies between magnon pairs by taking the expectation values of these terms with respect to two-magnon states in which the magnons have momenta $\pm k_l$.  For $J' > 2(J_1 + 4 J_2)$, where $k_l = 0$, Goldstone's theorem prevents the $\tilde{b}$ magnons from interacting.  As we lower $J'$ below $2(J_1 + 4 J_2)$, the $\tilde{b}$ magnons begin repelling increasingly strongly until the interaction energy diverges to $+ \infty$ at the stabilization curve \cite{app}.

At the upper critical field, we consider the quartic interaction terms containing $\tilde{c}^\dag(\pm k_u) \tilde{c}^\dag(\pm k_u) \tilde{c}(\pm k_u) \tilde{c}(\pm k_u)$.  For $J' > J_1 + 4 J_2$, where $k_u = 0$, the $\tilde{c}$ magnons repel with energy $4 J'$.  As we lower $J'$ below $J_1 + 4 J_2$, the $\tilde{c}$ magnons' repulsion initially continues decreasing but then increases strongly and again diverges to $+ \infty$ at the stabilization curve \cite{app}.

Therefore, away from the stabilization curve, the magnons do not form bound states (to this order in perturbation theory); single magnons condense as usual and the spins order perpendicular to the applied field.

\emph{Near the stabilization curve.}  Magnon interactions are much more complicated to treat near the stabilization curve, because both the $\tilde{b}$ and the $\tilde{c}$ magnons are near-gapless so we must consider interactions between them.  But the species carry opposite spin $S_z = \pm 1$, so the $U(1)$ spin symmetry ensures that these interactions preserve $N^{\tilde{b}} - N^{\tilde{c}}$ and total magnon number is not conserved.  Moreover, the Bogoliubov transformation breaks down as the coherence factors diverge.  This is because at the stabilization curve, the classical energy bands form tilted \emph{cones} rather than parabolas at the critical momenta $(\pm k_c, 0)$, with the $\epsilon_{\tilde{b}}(k)$ (i.e. $S_z = -1$) cones tilting away from $k = 0$ and the $\epsilon_{\tilde{c}}(k)$ (i.e. $S_z = +1$) cones tilting toward $k = 0$.  ($k_c = 2 \Delta \phi_0 / a > 0$, where $\Delta \phi_0$ is given by \eqref{stabilizationk0}.)

Instead, we return to the original sublattice magnons and coarse-grain them into a continuum coherent-state path integral using the precedure summarized in Table~\ref{Coarse-graining}.  In step 1 we transform the quadratic Hamiltonian to $H^{(2)} = \sum_k \hat{\Psi}^\dag \hat{M}(k) \hat{\Psi}(k)$, where the real symmetric matrix $\hat{M}(k)$ is defined in \eqref{Mhat}.  We found in Sec.~\ref{Spin waves} that the mode $\tilde{a}$ given by \eqref{atilde} is gapped for $k_y = 0$, so we project it out in step 2 and keep the low-energy degrees of freedom
\beq \label{Omega}
\Omega(k) := \left( \begin{array}{c} d(k, 0) \\ \hat{c}^\dag(-k, 0) \end{array} \right) := \left( \begin{array}{c} \big( a(k) - \hat{b}(k) \big) / \sqrt{2} \\ \hat{c}^\dag(-k, 0) \end{array} \right)
\eeq
with the effective Hamiltonian
\beq \label{H2eff}
H^{(2)}_\text{eff} = \sum_k \Omega^\dag(k)\, A(k)\, \Omega(k)
\eeq
where the matrix $A(k)$ is given in \eqref{A} for $k_y = 0$.  If we restrict the ``metric'' $g$ to the $2 \times 2$ matrix $\diag(1, -1)$, then the non-Hermitian matrix $g A(k_x)$ becomes defective at $(\pm k_c, 0)$ on the stabilization curve, which is why the Bogoliubov transformation breaks down.

\begin{table}
\caption{Algorithm for coarse-graining microscopic magnon operators into continuum fields.} \label{Coarse-graining}
\begin{enumerate}
\item Transform the magnon ladder operators to $\hat{\Psi}(k) := \left( a(k) \ \ \hat{b}(k) \ \ \hat{c}^\dag(-k) \right)^T$, where $\hat{b}(k) := e^{-\frac{1}{2} i k_x} b(k)$ and $\hat{c}(k) := e^{-\frac{1}{2} i k_1} c(k)$.
\item Transform $a(k)$ and $\hat{b}(k)$ to $\tilde{a}(k)$ and $d(k)$ defined by $\left( a(k) \mp \hat{b}(k) \right) / \sqrt{2}$, and drop terms involving $\tilde{a}(k)$.
\item Normal-order the $\Omega(k)$ operators defined in \eqref{Omega} and coarse-grain them to continuum operators.
\item Transform $\Omega(k, \tau)$ to $\Phi(k, \tau) := \big( \phi(k, \tau) \ \ \eta(k, \tau) \big)^T = U^\dag(k) \Omega(k, \tau)$, where $U$ is given by \eqref{U}.
\item Expand $\Phi(k, \tau)$ about the critical momenta $(\pm k_c, 0)$.
\item Remove terms involving the gapped $\eta_\alpha$ modes.
\item Rescale $\phi \to \varphi := \phi / \sqrt{S \Delta_\eta}$.
\end{enumerate}
\end{table}

In step 3 we coarse-grain and add imaginary time derivative terms $\int d\tau \left( d^\dag \partial_\tau d + \hat{c}^\dag \partial_\tau \hat{c} \right)$ to get the action
\beq \label{SfreeOmega}
S^{(2)} = \sum_{\alpha = R, L} \int \frac{d^2k}{(2 \pi)^2} d\tau \left( \Omega^\dag g \partial_\tau \Omega + S\, \Omega^\dag A \Omega \right).
\eeq
At the critical momenta,
\beq
A(\pm k_c, 0) \propto \left(
\begin{array}{cc}
1 & 1 \\
1 & 1
\end{array}
\right),
\eeq
and in step 4 we diagonalize it via the unitary transformation
\beq \label{U}
U(\pm k_c, 0) = \frac{1}{\sqrt{2}} \left(
\begin{array}{cc}
1 & 1 \\
-1 & 1
\end{array}
\right).
\eeq
We find that one eigenmode $\phi(k, \tau)$ becomes gapless at the critical momenta, while the other eigenmode $\eta(k, \tau)$ has a strictly positive gap $S \Delta_\eta = 2(4 S J' - h_c)$ \cite{app}.

In step 5 we expand $A(k)$ to second order and $U^\dag(k) g U(k)$ to first order about $(\pm k_c, 0)$.  For any field or function $f(k)$, we define $f_R(k) := f(k + (k_c, 0))$ and $f_L(k) := f(k - (k_c, 0))$.  The action becomes
\begin{align}
S^{(2)} &\approx \sum_{\alpha = R, L} \int \frac{d^2k}{(2 \pi)^2} d\tau \left[ \phi_\alpha^\dag \partial_\tau \eta_\alpha + \eta_\alpha^\dag \partial_\tau \phi_\alpha \pm \lambda k_x \phi_\alpha^\dag \partial_\tau \phi_\alpha \right. \nonumber \\
& \hspace{48pt} \left. + \phi_\alpha^\dag \left( S \kappa_x k_x^2 + S \kappa_y k_y^2 \right) \phi_\alpha + S \Delta_\eta\, \eta_\alpha^\dag \eta_\alpha \right],
\end{align}
where we have dropped kinetic terms for $\eta$ because it is gapped.  The constants $\kappa_x$, $\kappa_y$, and $\lambda$ are calculated in App.~\ref{Interactions}.

Integrating out the gapped modes $\eta_\alpha$ in step 6 gives
\begin{align}
S^{(2)} &= \sum_{\alpha = R, L} \int \frac{d^2k}{(2 \pi)^2} d\tau \left[ \vphantom{\frac{1}{S \Delta_\eta}} \phi_\alpha^\dag \left( S \kappa_x k_x^2 + S \kappa_y k_y^2 \pm \lambda k_x \partial_\tau \right) \phi_\alpha \right. \nonumber \\
&\hspace{93pt} \left. + \frac{1}{S \Delta_\eta} \partial_\tau \phi_\alpha^\dag \partial_\tau \phi_\alpha \right].
\end{align}
The $\lambda k_x \partial_\tau$ term (which arises because the cones in the classical energy bands are tilted) is small, because $\lambda \ll 1$ and it is a factor of $S$ smaller than the $k^2$ terms.  Moreover, as we demonstrate below, the region of interest does not flow to a fixed point under the renormalization group, so this term probably does not qualitatively affect the physics, and we neglect it.  The rescaling in step 7 finally gives
\beq
S^{(2)} = \sum_{\alpha = R, L} \int d^2x\, d\tau \left[ \varphi_\alpha^\dag \left( -c_x^2 \partial_x^2 - c_y^2 \partial_y^2 - \partial_\tau^2 \right) \varphi_\alpha \right].
\eeq
The theory has emergent Lorentz invariance with a slightly anisotropic speed of light $c_i := S \sqrt{\Delta_\eta \kappa_i}$.  We define $\partial_\mu := (\partial_\tau, c_x \partial_x, c_y \partial_y)$ and set $c_x = 1$.  (Since $c \propto S$, $S$ is no longer a large parameter under this choice of normalization.)

\emph{Quartic interactions near the stabilization curve.}  Applying the same procedure to the quartic terms $H^{(4)}$ gives the complete action
\begin{align}
&S \left[ \varphi_R^\dag, \varphi_R, \varphi_L^\dag, \varphi_L \right] = \int d^2x\, d\tau \label{SPhi} \\
&\left[ \sum_{\alpha = R, L} \left( \partial_\mu \varphi_\alpha^\dag \partial_\mu \varphi_\alpha + \frac{1}{6} V_p \left( \varphi_\alpha^\dag \varphi_\alpha \right)^2 \right) + \frac{1}{3} V_a\, \varphi_R^\dag \varphi_R\, \varphi_L^\dag \varphi_L \right], \nonumber
\end{align}
where the parallel and antiparallel coupling constants $V_p$ and $V_a$ are defined in \eqref{Vs} (the reason for their nonstandard normalization will become clear in Sec.~\ref{RG}).  Both couplings are very close to proportional to $J'$; the ratio $V_a / V_p$ begins at $5.32$ at the Lifshitz point and increases with $J'$, quickly saturating to $6.85$.

The linearly dispersing $c$ magnons with spin $+1$ and $b$ magnons with spin $-1$ can be considered analogous to particles and holes in a fermionic system with two Dirac cones, with the field $h_c$ corresponding to a chemical potential tuned to half-filling, making both the emergent Lorentz invariance and the conservation of ``charge'' $N^b - N^c$ very natural.  The emergence of Lorentz invariance near the stabilization curve also explains the divergence of the Bogoliubov transformation and magnon self-energies calculated in Section~\ref{Corrections}: we were treating the magnon interactions perturbatively, but relativistic scalar $\varphi^4$ couplings have positive mass dimension in $d = 3$ spacetime dimensions, so we expect weak bare interactions to be relevant under the renormalization group and flow to strong coupling at low energy, causing our perturbative treatment to break down.

\emph{Mean-field analysis of action}.  Action \eqref{SPhi} clearly has two independent global $U(1)$ symmetries $\varphi_\alpha \to e^{i \theta_\alpha} \varphi_\alpha$, two $\mathbb{Z}_2$ symmetries $\varphi_\alpha \leftrightarrow \varphi_\alpha^\dag$, and a third ``chiral'' $\mathbb{Z}_2$ symmetry $\varphi_R \leftrightarrow \varphi_L$.  If we add a small mass term $r \left( \varphi^\dag_R \varphi_R + \varphi^\dag_L \varphi_L \right)$ and make a Landau-Ginzburg mean-field ansatz $\langle \varphi_\alpha \rangle = \sqrt{\rho_\alpha} e^{i \theta_\alpha}$, the mean-field energy density becomes
\beq
\mathcal{H}_\text{MF} = r \left( \rho_R + \rho_L \right) + \frac{1}{6} V_p \left( \rho_R + \rho_L \right)^2 + \frac{1}{3} (V_a - V_p)\, \rho_R\, \rho_L.
\eeq
If $V_a > V_p$ (which is true along the entire stabilization curve), then as we tune $r$ negative, the mean-field energy is minimized when one of the fields (we will assume $\varphi_R$) acquires an expectation value with $\rho_R = 3 |r| / V_p$ and the other does not \cite{Liu}.  Therefore the chiral $\mathbb{Z}_2$ and only one of the two $\mathbb{Z}_2 \times U(1)$ symmetries are broken.

The procedure in Table~\ref{Coarse-graining} coarse-grains the microscopic spin operator $S_r^+$ to 
\beq \label{Splus}
\langle S_r^+ \rangle = A S e^{i (k_c x + \theta_R)},
\eeq
where $A = a \sqrt{\Delta_\eta \rho_R/2}$ on the chains and $A = -a \sqrt{\Delta_\eta \rho_R}$ on the interstitial spins \cite{app}.  This state corresponds to the semiclassical cone state illustrated in Fig.~\ref{Cone}.  The coarse-grained spin Hamiltonian has two (locally) $U(1)$ symmetries -- translational invariance and spin rotation about the applied field -- which are broken individually, but there is a combined $U(1)$ symmetry consisting of translation by an arbitrary $\Delta x$ and spin-space rotation by $k_c\, \Delta x$, corresponding to the action's symmetry $\varphi_L \to \varphi_L e^{i \theta_L}$, which remains unbroken.

\emph{Interactions at the saturation field}.  At the saturation field, the condensed $\tilde{c}$ magnons always repel, so there is no pair condensation and magnon interactions do not change the classical saturation field at this order (App.~\ref{Corrections app}).  The $\tilde{a}$ magnon, which lives only on the chains, is always gapped at the classical saturation field for $J' > 0$, but the gap vanishes as $J' \rightarrow 0$ and the chains decouple (and the $\tilde{a}$ and $\tilde{c}$ magnons become equivalent).  Moreover, these magnons attract at most momenta, suggesting that for small $J'$ and $S$, the leading instability at the saturation field may be a spin nematic phase with condensed bound pairs of $\tilde{a}$ magnons.  (This is known to be the case for $J' = 0$ and $S = 1/2$.)

\section{RG analysis of action near stabilization curve \label{RG}}

It will be convenient to rewrite $\varphi_\alpha = \frac{1}{\sqrt{2}} \left( \varphi_{\alpha,1} + i \varphi_{\alpha,2}\right)$ in action \eqref{SPhi} in terms of its real components $\vec{\varphi}_\alpha := (\varphi_{\alpha,1}, \varphi_{\alpha,2})$:
\begin{align}
S \left[ \vec{\varphi}_R, \vec{\varphi}_L \right] = \int d^3x &\left[ \sum_{\alpha = R, L} \bigg( \frac{1}{2} (\partial_\mu \vec{\varphi}_\alpha)^2 + \frac{1}{2}r\, \vec{\varphi}_\alpha^2 \right. 
\label{SVecPhi} \\
&\ \left. \vphantom{\sum_\alpha} + \frac{1}{4!} V_p \left( \vec{\varphi}_\alpha^2 \right)^2 \bigg) + \frac{2}{4!} V_a \vec{\varphi}_R^2 \vec{\varphi}_L^2 \right], \nonumber
\end{align}
whose symmetry group is now thought of as $O(2) \times O(2) \times \mathbb{Z}_2$.  Refs.~\onlinecite{Kosterlitz, Calabrese03, Calabrese02} studied the RG flow of this action and found three fixed points.  The first is the Gaussian fixed point $V_p = V_a = 0$, which is unstable.  The second, ``decoupled fixed point'' has $V_a = 0,\ V_p \neq 0$ and corresponds to a tetracritical intersection of two $O(2)$ critical lines.  This fixed point can be shown nonperturbatively to be stable.  At the third fixed point, $V_p = V_a \neq 0$ and the symmetry group is enlarged to $O(4)$, because the action becomes invariant under any transformation that preserves the norm of $\vec{\varphi} := \vec{\varphi}_R \oplus \vec{\varphi}_L$ (so $\vec{\varphi}^2 = \vec{\varphi}_R^2 + \vec{\varphi}_L^2$).  This fixed point has been shown to be unstable in $d = 3$ by a five-loop $\epsilon$-expansion \cite{Calabrese03} and by a six-loop fixed-dimension perturbative expansion \cite{Calabrese02}.  (In the mean-field analysis of Sec.~\ref{Interactions}, this fixed point separated the phases in which the chiral $\mathbb{Z}_2$ symmetry was broken and unbroken.)  The bare couplings of our model lie on the $V_a > V_p$ side of the $O(4)$ fixed point, so we expect $V_a$ to flow to stronger coupling than $V_p$.  To our knowledge, the nature of the phase transition outside of the decoupled fixed point's basin of attraction has not been studied in detail.

\emph{Chiral liquid phase near stabilization curve.}  We now argue that near the stabilization curve, quantum fluctuations that are not captured by mean-field theory create a strongly fluctuating, short-range-entangled ``chiral liquid'' phase with no classical analogue in between the magnetization plateau and the semiclassical cone state, in which the chiral $\mathbb{Z}_2$ symmetry is broken but the $U(1)$ symmetries of translation and spin rotation both remain unbroken.  As in our mean-field analysis, we rewrite \eqref{SVecPhi} by separating out the $O(4)$ symmetric terms:
\begin{align}
&S = \int d^3x \bigg[ \frac{1}{2} (\partial_\mu \vec{\varphi})^2 + \frac{1}{2}r\, \vec{\varphi}^2 + \frac{V_a + V_p}{2 \times 4!} \left( \vec{\varphi}^2 \right)^2 \nonumber \\
&\hspace{50pt} - \frac{V_a - V_p}{2 \times 4!} \left( \vec{\varphi}_R^2 - \vec{\varphi}_L^2 \right)^2 \bigg] \\
&\hspace{4pt}:= S_{O(4)} - \int d^3x \left[ \frac{V_a - V_p}{2 \times 4!} O(x)^2 \right], \nonumber
\end{align}
where the $O(4)$-symmetry-breaking order parameter $O(x) := \vec{\varphi}_R(x)^2 - \vec{\varphi}_L(x)^2$.  We then introduce an auxiliary Hubbard-Statonovich real scalar field $\chi$ that couples to the chiral order parameter:
\beq \label{S}
S = S_{O(4)} + \int d^3x \left( \frac{1}{2} \frac{4!}{V_a - V_p} g^2 \chi^2 - g \chi O \right),
\eeq
where the dimensionful coupling constant $g$ fixes the normalization of the $\chi$ field.  Separating out the free action $S_{0,\chi} := \int d^3x\, \frac{1}{2} \frac{4! g^2}{V_a - V_p} \chi^2$ and integrating over the $\vec{\varphi}_\alpha$ fields gives
\beq \label{Z}
Z = \int D\chi\, e^{-(S_{0,\chi} + \Delta S)},
\eeq
where
\beq \label{DeltaSDef}
\Delta S[\chi] := -\ln \left[ \left \langle \exp \left[ \int d^3x\, (g \chi(x) O(x)) \right] \right \rangle_{S_{O(4)}} \right].
\eeq

Near the $SO(4)$ point, this action simplifies to \cite{app}
\beq \label{DeltaS}
\Delta S \approx \int d^3x\, d^3 x'\, \left( -g^2 \chi(x) \chi(x') \langle O(x) O(x') \rangle_{S_{O(4)}} \right).
\eeq
Away from the critical point $r = 0$, the correlation function takes the ``Ornstein-Zernike'' form \cite{Kennedy, Rams}
\beq
\langle O(x) O(x') \rangle_{S_{O(4)}} \approx \frac{C\, f\left( |x - x'|/\xi \right)}{|x - x'|^{2 \Delta}},
\eeq
where the order parameter $O$ has correlation length $\xi$ and scaling dimension $\Delta$, $C$ is a nonuniversal dimensionful constant, and $f$ is a universal dimensionless function.  The Gaussian theory has scaling dimension $1$, but near the $O(4)$ point the interactions renormalize the scaling dimension to $\Delta = 1.19$ \cite{app}.  Fourier transforming $\chi(x)$ and rescaling to $y := x/\xi$,
\begin{align}
&\Delta S = \label{DeltaSFT} \\
&\ \, \int \frac{d^3k}{(2 \pi)^3} \left[ -g^2 |\chi(k)|^2\ C \xi^{3 - 2 \Delta} \int d^3y \left( e^{-i \xi k y} \frac{f(|y|)}{|y|^{2 \Delta}} \right) \right]. \nonumber
\end{align}
$f(|y|)$ decays exponentially at large $y$, so the Fourier transform is analytic in a neighborhood of $k = 0$ and we can Taylor expand it.  We find that setting $g^2 = \xi^{2 \Delta - 5}/(b C)$ (where the universal constant $b$ is defined in \eqref{b}) normalizes $\chi$ appropriately and gives
\beq \label{DeltaSNormalized}
\Delta S = \int d^3 x \left[ \frac{1}{2} \left( \partial_\mu \chi \right)^2 - \frac{1}{2} \frac{2}{b\, \xi^2} \chi^2 \right],
\eeq
so the coupling to $O$ makes the $\chi$ field dynamical and also renormalizes its mass downward.

\begin{figure}
\includegraphics[width=\columnwidth]{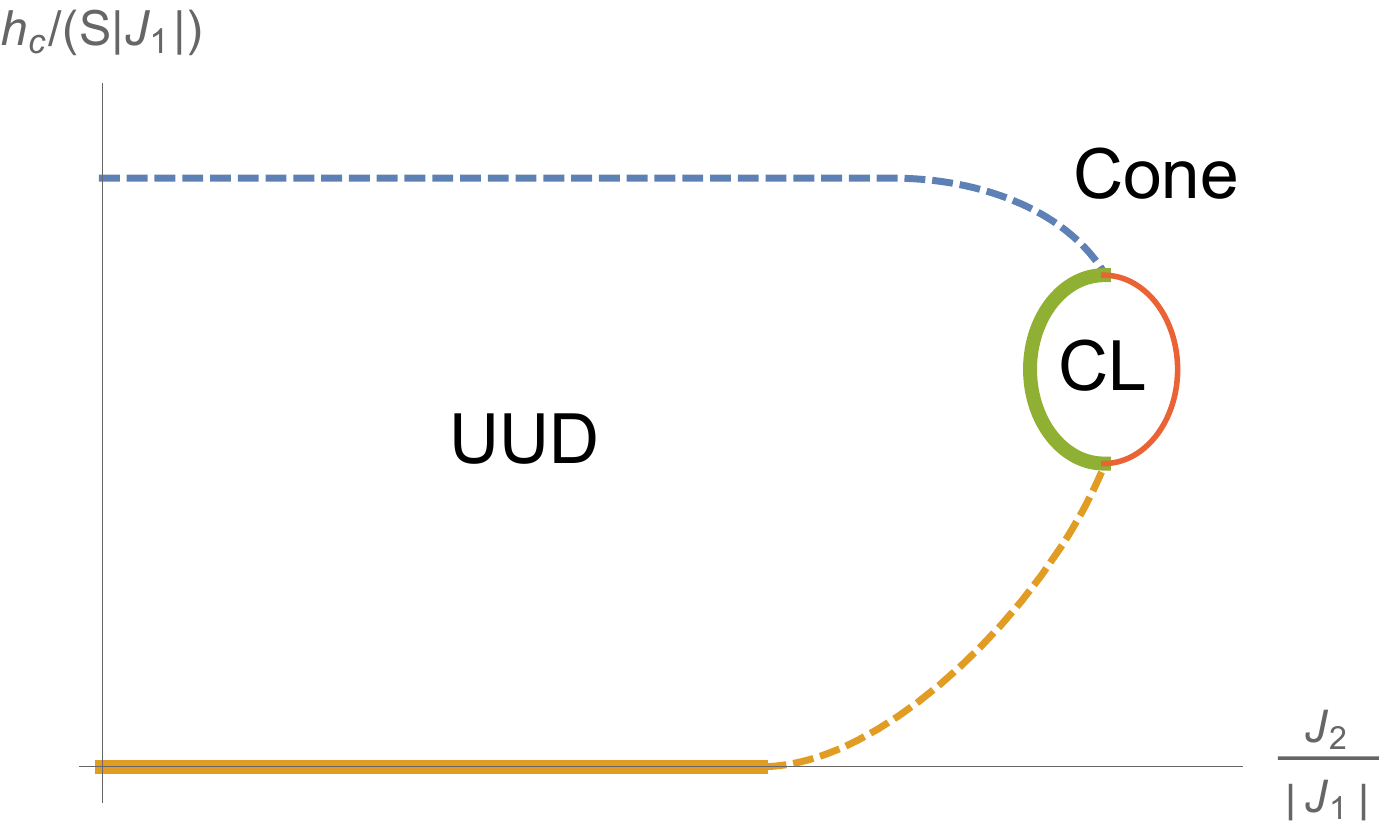}
\caption{Schematic quantum phase diagram at constant $J'$.  The UUD state breaks no symmetries, the gapped chiral liquid (CL) phase only breaks chiral symmetry, and the gapless cone state breaks both chiral and a $U(1)$ symmetry combining translation and spin rotation.  The thick solid lines and dashed lines represent first- and second-order transitions respectively.  We did not investigate the nature of the transition between the chiral liquid and cone phases represented by the red line.  In a 3D phase diagram like that of Fig.~\ref{3D LT phase diagram} that includes the applied field, the chiral liquid phase would appear as a thin tube around the stabilization curve where the two sheets meet.} \label{CL phase diagram}
\end{figure}

The complete action in \eqref{Z} becomes
\beq
S[\chi] = \int d^3 x \left[ \frac{1}{2} \left( \partial_\mu \chi \right)^2 + \frac{1}{2} \left( \frac{4!\, \xi^{2 \Delta - 3}}{(V_a - V_p) C} - 2 \right) \frac{1}{b\, \xi^2} \chi^2 \right].
\eeq
If the $O(4)$-symmetric interaction renormalized $\Delta$ to a value higher than $3/2$, then $\chi's$ renormalized mass would remain positive as $\vec{\varphi}_\alpha$'s mass $r \to 0$ and $\xi$ diverged, and $\chi$ would not acquire an expectation value anywhere inside the classical plateau.  But $\Delta = 1.19 < 3/2$, so chiral symmetry is broken slightly inside the plateau, at the couplings at which $\vec{\varphi}_\alpha$'s mass remains positive but becomes small enough that
\beq
\xi > \left( 2 C \frac{V_a - V_p}{4!} \right)^{-\frac{1}{3 - 2 \Delta}}.
\eeq
Fig.~\ref{CL phase diagram} displays a schematic phase diagram.

\emph{Nature of phase transition to chiral liquid.}  Ref.~\onlinecite{Calabrese03} argues on general principles \cite{Mukamel, Domany} that for bare couplings that lie outside the decoupled fixed point's RG basin of attraction, phase transitions (in our case, between the plateau and the chiral liquid phase) are probably first-order, although the authors do not perform a calculation.  We present a calculation for our model that strengthens their conclusion, although we make one uncontrolled approximation.  We determine the order of the transition by calculating the sign of the $\chi^4$ coupling.  The $O(4)$-symmetric $\vec{\varphi}^4$ interaction makes this quite difficult, so we will neglect it and only consider the cubic interaction $g \chi O$.  

In this approximation, \eqref{DeltaSDef} becomes $e^{-\Delta S[\chi]} \propto \int D\vec{\varphi}_R\, D\vec{\varphi}_L\, e^{-\tilde{S}}$, where
\beq \label{Stilde}
\tilde{S}[\vec{\varphi}_\alpha, \chi] := \int d^3x \left[ \frac{1}{2} (\partial_\mu \vec{\varphi})^2 + \frac{1}{2}r\, \vec{\varphi}^2 - g \chi O \right].
\eeq
Defining $P := -\partial^2 + r$, $K(x) := 2 g \chi(x)$ and integrating over $\vec{\varphi}_\alpha$ gives \cite{app}
\beq \label{DeltaSTrace}
\Delta S = -\frac{1}{2} \sum_{n=1}^\infty \frac{1}{n} \Tr \left[ \left( P^{-1} K \right)^{2n} \right].
\eeq
In the momentum basis, $P^{-1} \to 1/(k^2 + r)$ is just the usual free Green's function for $\vec{\varphi}_\alpha$, but in general $K$ is not translationally invariant so $P^{-1}$ and $K$ do not commute, making the action difficult to work with.  But we only need the potential energy function $U(\chi)$, so we can take $\chi$ to be constant, and the action $S[\chi(x)]$ simplifies to $S(\chi) = V U(\chi)$ (where $V$ is the volume of spacetime), and \cite{app}
\begin{align}
U(\chi) &= -\frac{1}{2} \sum_{n=1}^\infty \frac{(2 g \chi)^{2n}}{n} \int \frac{d^3k}{(2 \pi)^3} \frac{1}{(k^2 + r)^{2n}} \label{Uchi} \\
&= -\frac{r^{3/2}}{\pi} \sum_{n=1}^\infty \frac{1}{n (4n - 3) 4^n} \binom{4 n-3}{2 n-1} \left( \frac{g \chi}{r} \right)^{2n}. \nonumber
\end{align}
The $n = 2$ term $-g^4/(16 \pi\, r^{5/2})\, \chi^4$ is negative, suggesting that the transition to the chiral liquid phase is first-order.

Of course, incorporating the $O(4)$-symmetric $\vec{\varphi}^4$ interaction may well change the sign of the $\chi^4$ coupling.  The interaction changes the universality class at $\chi = 0$, so its effect may well be important but is hard to evaluate perturbatively.  For example, \eqref{Uchi} seems to indicate that \emph{all} the derivatives of $U(\chi)$ at $\chi = 0$ are negative, suggesting that $U(\chi)$ has no stable stationary points.  However, the large $\chi$ regime is presumably stabilized by the $\vec{\varphi}^4$ term and is anyway outside the regime of the perturbative expansion.  Beyond perturbation theory, we can say based on the usual grounds of universality that, {\em if} the transition from the UUD state to the chiral one becomes continuous, the critical behavior should be in the Ising universality class.  This is an entirely consistent, if not perturbatively accessible, possibility.

\section{Chiral liquid phase \label{Wavefunction}}

In this section we discuss the physical properties and symmetry of the chiral liquid state.  First, we make a simple observation.  Because it corresponds to a {\em neutral} condensate with net $S^z=0$, the chiral liquid does not break the $U(1)$ spin rotation symmetry.  Consequently, the magnetization remains quantized and equal to the value of $1/3$ of saturation: {\em the chiral liquid phase is part of the plateau}.

Now we explore the chiral order parameter in physical terms.  The microscopic order parameter for the chiral-liquid phase must respect the $U(1)$ spin and translational symmetries, but violates the Hamiltonian's chiral (i.e. inversion) symmetry.  A natural candidate is the component $J_{ij}^z = \frac{1}{2} i (S_i^+ S_j^- - S_i^- S_j^+)$ of the ``spin-current'' operator $\bm{J}_{ij} := \bm{S}_i \times \bm{S}_j$.  The spin current can be interpreted semiclassically as describing the precession of each spin due to the effective field from the other spin \cite{Cheong}.  The usual physical origin of spin current is a Dzyaloshinskii-Moriya (DM) coupling $H_\text{DM} = \sum_{i,j} -\bm{D}_{ij} \cdot \bm{J}_{ij}$, where the Dzyaloshinskii vector $\bm{D}_{ij}$ is induced by spin-orbit coupling \cite{Dzyaloshinskii, Moriya}.

We see from \eqref{S} that a saddle-point expansion of $\chi$ about its nonzero expectation value gives that $\langle \chi(x) \rangle = (V_a - V_p) / (4! g) \langle O(x) \rangle$, so we need to coarse-grain $J^z(\Delta r)$ in terms of $O(x)$.  ($O$ is quadratic in magnon operators, but we cannot perform a mean-field decomposition like $\langle a^\dag a \rangle \sim \langle a^\dag \rangle \langle a \rangle$ as in Sec.~\ref{Interactions}, because the unbroken $U(1)$ spin symmety prevents single-magnon condensation.)  The unbroken translational symmetry requires that $\langle J^z_{rr'} \rangle$ only depend on $r - r' := \Delta r = (\Delta x, \Delta y)$.  The procedure in Table~\ref{Coarse-graining} gives \cite{app}
\begin{align}
\langle J^z_{(ss')}(\Delta r) \rangle &= \frac{1}{4} c_{ss'} S^2 a^2 \Delta_\eta \langle O(x) \rangle \sin(k_c \Delta x) \nonumber \\
&= c_{ss'} J_0\, \langle \chi(x) \rangle \sin(k_c \Delta x), \label{Jz}
\end{align}
where $J_0 := 6 S^2 a^2 g \Delta_\eta / (V_a - V_p)$ and $s, s'$ represent sublattice indices.  $c_{ss'} = 1$ for $s, s' \in \{ A, B \}$, $c_{CC} = 2$, and $c_{sC} = -\sqrt{2}$ for $s \in \{ A, B \}$.  This expression holds over distances $\Delta r$ much shorter than the connected correlation length for $\chi$.

\begin{figure}
\includegraphics[page=3,width=\columnwidth]{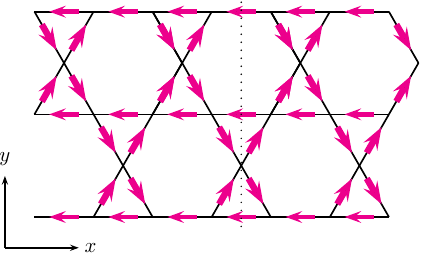}
\caption{Spin-current configuration in the chiral liquid phase.  The magenta arrows indicating the direction of spin current flow; in the orthogonal ground state the flow is reversed.  The spin current on the diagonal bonds, represented by thicker arrows, is larger by a factor of $\sqrt{2}$ and determines the net current flow.  The ground state is chiral and spontaneously breaks the lattice symmetry of reflection about the dotted line.} \label{Spin currents}
\end{figure}

The spin currents are illustrated in Fig.~\ref{Spin currents}.  The spin current on the chains and the spin current between the chains flow in opposite directions, and the triangles have a net circulation but the hexagons do not.  The magnetization from the applied field results in a nonzero scalar chirality on the triangles.  If we define the scalar chirality to be $\langle {\bf S}_A \cdot ({\bf S}_B \times {\bf S}_C) \rangle$ on every triangle (the convention used in Ref. \onlinecite{ChubukovStarykh}), then it is uniform across all triangles in the ground state, and the order-parameter field $\chi(x)$ is proportional to the coarse-grained scalar chirality.  If we instead define the scalar chirality to be $\langle {\bf S}_A \cdot ({\bf S}_B \times {\bf S}_C) \rangle$ on the upward-pointing triangles and $\langle -{\bf S}_A \cdot ({\bf S}_B \times {\bf S}_C) \rangle$ on the downward-pointing triangles (the convention used in Ref. \onlinecite{Zhu}), then the scalar chirality has one sign on the upward-pointing triangles and the opposite sign on the downward-pointing ones, and the order parameter $\chi(x)$ is proportional to the coarse-grained \emph{staggered} scalar chirality.

We argued above that the symmetries of the chiral liquid phase alone require a rather exotic composite ``bond-nematic'' order parameter, because the unbroken $U(1)$ spin symmetry prevents single-spin ordering, and the unbroken translational symmetry prevents any kind of valence-bond crystal or spin-density-wave ordering.  As in the case of the frustrated $J_1$-$J_2$ ferromagnetic chain, we can interpret this ordering in terms of pairs of weakly attracting magnons.  If we start with the applied field inside the plateau, magnons are completely gapped.  As we increase $J_2$, the attractive interaction increases until it becomes large enough to close the gap to two-magnon condensation in the chiral liquid phase, while single magnons remained gapped.  As we increase $J_2$ still further, the single-magnon gap closes as well and the pair condensation is preempted by ordinary superfluid condensation.  But unlike in the case of $J_1$-$J_2$ chain magnon pair-condensation, which breaks the $U(1)$ spin symmetry down to $\mathbb{Z}_2$ with a order parameter of the form $\langle S_i^- S_{i+1}^- \rangle$, in our case the attracting magnons have opposite spins, so their pair-condensation with order parameter $\langle i S_i^+ S_{i+1}^- + \text{h.c.} \rangle$ leaves the $U(1)$ symmetry completely unbroken.

This pair condensation is somewhat reminiscent of Cooper pairing in BCS superconductivity, but with an important difference.  The microscopic degrees of freedom in a superconductor are fermionic, so Cooper-paired electrons must either be in a spin singlet with a spatially even order parameter (as in $s$-wave pairing) or in a spin triplet with spatially odd order parameter (as in $p$-wave pairing).  But magnons are bosonic, and in our case the bound pairs' spin and orbital degrees of freedom are \emph{both} odd, resulting in an overall symmetric wavefunction. 

\begin{figure}
\includegraphics[page=4,width=\columnwidth]{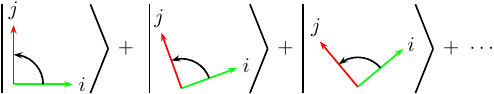}
\caption{Schematic semiclassical illustration of a bond connecting sites $i$ and $j$ with positive spin current $J_{ij}^z$.  The page represents the $S^x$-$S^y$ plane in spin space, with $S^z$ pointing outward.  The green and red arrows represent the spins at sites $i$ and $j$ respectively.  The sum represents an equal-weighted integral over all orientations in the $S^x$-$S^y$ plane.} \label{Semiclassical spin current}
\end{figure}

We can gain some intuition relating the chiral liquid state to the semiclassical cone state by considering the eigenstate with $J_{ij}^z = +S$ schematically illustrated in Fig.~\ref{Semiclassical spin current}.  A single spin-current bond can be thought of as a uniform superposition of each pair of consecutive spins in a classical spiral state: the chirality of the rotation is preserved but the dipole moment of each spin averages to zero.  (This is analagous to a time-averaged, circularly polarized classical electromagnetic wave, whose electric and magnetic fields average to zero while their cross product does not.)  As we tune $J_2$ from the cone into the chiral liquid phase, the system loses its translational order but preserves its ``memory'' of the sense in which the cone was rotating.  (However, an actual superposition of all possible translations of a classical cone state is an unstable cat state.)

\emph{Experimental signatures of spin-current order.}  The signatures are largely similar to those predicted for the chiral spin current state in Ref.~\onlinecite{ChubukovStarykh}.  The spin current $\bm{J}$ is even under time reversal and odd under parity inversion, just like electric polarization, so it is natural for them to couple linearly.  This results in a magnetoelectric effect in which the spin-wave dispersions depend linearly on an applied \emph{electric} field.  The applied magnetic field $\bm$ at a bond $\bm{e}_{ij}$ connecting sites $i$ and $j$ also induces a local electric polarization $\bm{P}_{ij} \parallel \bm{J}_{ij} \times \bm{e}_{ij}$ \cite{Cheong, Katsura, Mostovoy}.  In the triangular-lattice case this electric polarization averages to zero over a magnetic unit cell, but in our case there is a net spin current along the $x$-axis, so the applied field induces an average electric polarization $\bar{\bm{P}} \parallel \bm{h} \times \hat{x}$.  Also, if we allow a small intersite hopping amplitude $t$, then the local electric polarizations induce \emph{charge} current circulation in opposite directions about the upward- and downward-pointing triangles.  This in turn induces weak staggered magnetic dipole moments at the \emph{centers} of the triangles, which could be measured by neutron scattering, nuclear magnetic resonance (NMR), or muon spectroscopy \cite{Al-Hassanieh}.  The combination of strong applied fields, DM couplings, and geometric frustration may also lead to a magnon anomalous thermal Hall effect, like the one recently measured in the pyrochlore magnet $\text{Lu}_2\text{V}_2\text{O}_7$ \cite{Onose}.

\section{Comparison with other models} \label{Triangular lattice}

A similar analysis has been performed on a model of an anisotropic nearest-neighbor antiferromagnet on a triangular lattice \cite{Alicea, ChubukovStarykh}; our results have some similarities but important differences.  The triangular model has a $1/3$-magnetization UUD state analagous to our kagom\'{e} plateau, which is frustrated by the anisoptropy instead of our $J_2$ coupling.  Both models have a lobe-shaped plateau phase adjacent to semiclassical noncoplanar chiral states.  At both models' ``stabilization curves'' (a single point in the triangular model) where the $S^z = +1$ and $S^z = -1$ magnons simultaneously condense, the coherence factors diverge and magnon pair-condensation preempts single-magnon condensation, resulting in a chiral spin-current phase (compare our Figs.~\ref{CL phase diagram} and \ref{Spin currents} to Figs.~1(c) and 2 of Ref.~\onlinecite{ChubukovStarykh}).

But unlike our model, the triangular model's plateau is classically unstable and requires quantum fluctuations for stability, and it spontaneously breaks translational symmetry by singling out one sublattice to align against the field.  The triangular model's classical ground state is nonchiral for all couplings, so the classical magnon instability occurs at $k_c = 0$, giving a small parameter about which to expand and allowing a more controlled analytic treatment than is possible for our model.  Also, the triangular chiral-liquid phase has equal spin current on every bond and thus no net spin current, unlike our model.

Another study considered a system equivalent to $S = 1/2$ spins on a kagom\'{e} lattice coupled by spatially isotropic antiferromagnetic first-, second-, and third-nearest-neighbor $XXZ$ couplings at one-third of the saturation magnetization \cite{Zhu}.  The authors found a bosonic fractional quantum Hall phase with semionic excitations, whose ground states are two sets of $\nu = 1/2$ Laughlin states related by time-reversal symmetry.  The system spontaneously breaks the time-reversal symmetry while the twofold topological degeneracy remains robust.  The time-reversal-symmetry breaking is extremely analagous to our system's chiral symmetry breaking, and the order parameter is again given by a circulating spin current $J^z$.  However, in the $XXZ$ system, there is no net spin current, and the current circulates in the same direction around the upward- and downward-pointing triangles, so there is a current flow around the hexagons (compare our Fig.~\ref{Spin currents} to Fig.~3 of Ref.~\onlinecite{Zhu}).  Like ours, the $XXZ$ model's chiral phase occurs near a frustration-induced phase transition of the corresponding classical model.

\section{Comparison with volborthite experimental data \label{Experiment}}

\begin{figure}
\includegraphics[width=\columnwidth]{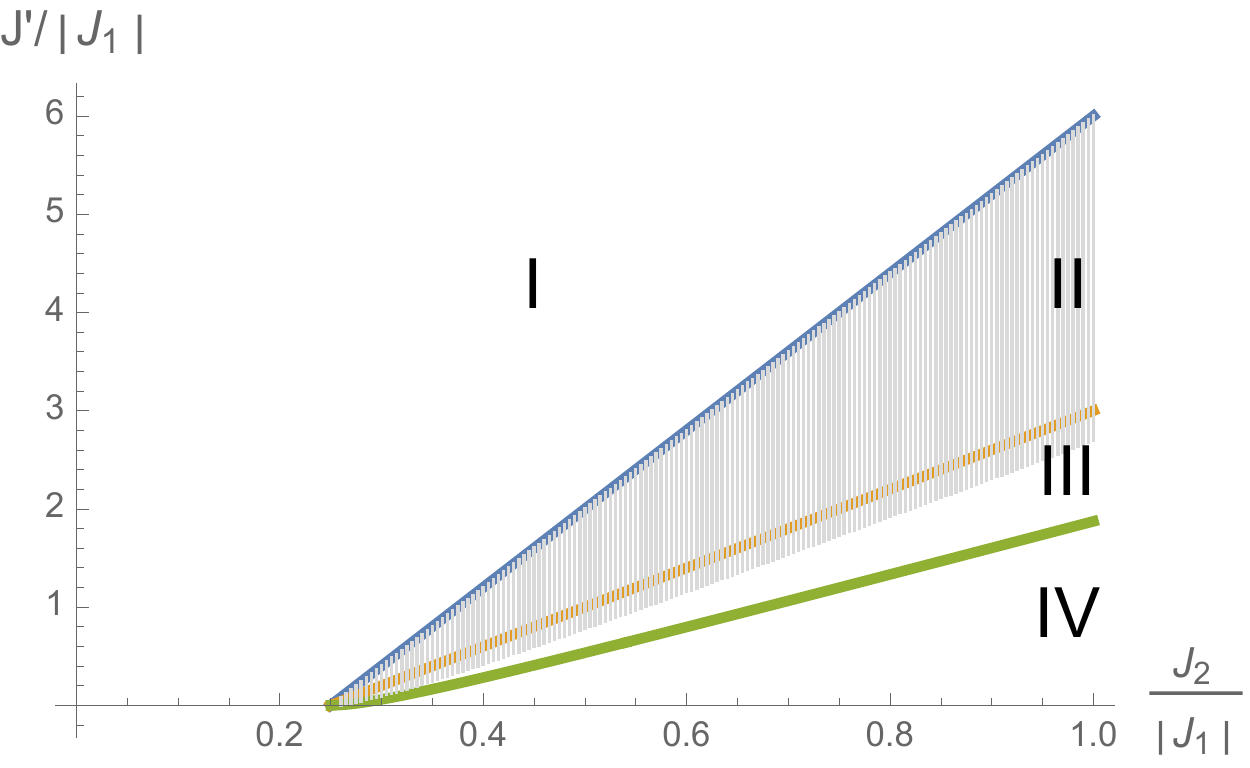}
\caption{The gray-shaded region of the classical phase diagram is the portion compatible with the experimental result $0 < h_l / h_u \leq 0.233$.  The labeled regions are the same as in Fig.~\ref{LT phase diagram}.} \label{Exp phase diagram}
\end{figure}

Here we consider on purely phenomenological grounds whether the model studied here may be relevant for volborthite.  On first sight, the presence of a broad plateau is encouraging.  Let us look at this more quantitatively.  Experiments indicate that the lower and upper critical fields of the magnetization plateau for volborthite are 28 T and at least 120 T \cite{Ishikawa, Yamashita}, so $h_l / h_u \leq 0.233$.  We show in \eqref{hlhu} that on the boundary between regions II and III, $h_l / h_u < (5 - \sqrt{17}) / 4 = 0.219$ only barely satisfies this constraint, so almost all of region III is experimentally excluded.  The classically non-excluded region is shown in Fig.~\ref{Exp phase diagram}.  If the plateau could be shown to extend to 128 T, then region III would be entirely excluded, and the model predicts that (up to quantum corrections) $h_u = 2 S J'$ and $h_\text{sat} = 6 S J'$, so a measurement of the upper critical field would allow a direct prediction of $J'$ and $h_\text{sat}$.  Unfortunately, the saturation field is far out of reach of currently achievable non-destructive measurements.

The result that quantum corrections broaden the plateau helps to reconcile two conflicting arguments for the relative importance of the $J'$ coupling: the qualitative agreement between experiment and the quantum phase diagram of the decoupled $S=1/2$ $J_1$-$J_2$ chains \cite{Ishikawa} suggests that the $J'$ coupling may not dominate the physics, but the classical analysis illustrated in Fig.~\ref{Exp phase diagram} suggests that the small value of $h_l / h_u$ requires a $J'$ coupling significantly larger than $J_1$ and $J_2$.  But the quantum fluctuations' broadening of the plateau increases the range of the phase diagram with a small ratio $h_l / h_u$, and therefore decreases the minimum $J'$ required to produce it.  So the actual value of $J'$ in volborthite may well be lower than the range indicated in Fig.~\ref{Exp phase diagram} by the classical analysis.  For example, the classical analysis indicates that $h_u = 2 S J' > 120$ T and so $J'$ is stronger than the Curie-Weiss temperature of -115 K \cite{Ishikawa}.  But the leading-ordering quantum fluctuations raise the upper critical field by about $60\%$ in this region, so a $J'$ of only 75 T (well below the Curie-Weiss temperature) may be enough to stabilize such a broad plateau (of course, the specific numerical values should not be taken too seriously).

From the above discussion we see that the plateau width and location could be reconciled within the frustrated chain model.  However, there is more difficulty in understanding the physics below the plateau.  For much of this field range, NMR experiments find an incommensurate but collinear spin-density-wave-like state \cite{Ishikawa}.  This is quite far from the semiclassical result, for which the region below the plateau is umbrella-like.  A further difficulty is the narrow regime just below but in the vicinity of the plateau onset, where some anomalies in the magnetization curve suggest either a narrow range of intermediate phase or possibly a two-phase coexistence regime associated with a first order transition (though Ref.~\onlinecite{Ishikawa} argues against the latter).  We considered a possible magnon-pairing instability to a $U(1)$-symmetry-breaking nematic state, but found that this was not present in the frustrated chain model in the relevant regime within the spin wave approach.  Furthermore, the calculations in Sec.~\ref{Interactions} indicate a continuous second order transition at the plateau edge.  

These considerations cast doubt on the validity of the frustrated ferrimagnetic chain model to volborthite.  Two possible reconciliations are: (1) effects beyond the semiclassical approximation reconcile the model with volborthite for $S=1/2$, or (2) another model is required.  Regarding possibility (1), the isolated frustrated ferromagnetic chain ($J'=0$) is known to show strong quantum effects well beyond the semiclassical description \cite{Arlego, Kolezhuk}.  It is not impossible that these occur here as well, although the breadth of the plateau seems to require rather large $J'$, which presumably suppresses some of these effects.  In support of possibility (2), more recent density functional calculations indicate a more complex spin Hamiltonian in which an appropriate picture is in terms of spin trimers \cite{Janson16}.  That model also features a plateau, but the low field properties might be in closer accord with experiment.  In particular, a $U(1)$-symmetry-breaking spin nematic state was argued to exist below the plateau, possibly describing the aforementioned anomalies there.  We tend to favor the simpler explanation that the spin trimer picture is closer to correct.

\section{Conclusion \label{Conclusion}}

We have found that the classical Hamiltonian \eqref{Hamiltonian} has a broad $1/3$ magnetization plateau for sufficiently large $J'$, surrounded at higher and lower fields by a precessing cone state.  We calculated the quantum corrections to the plateau's lower and upper critical fields from the magnon self-energies and found a broadening of the plateau in both directions (to leading order in $1/S$).  The shifts in the lower and upper critical fields relative to their classical values are about $-0.73 / S$ and $+0.29 / S$, respectively.  We found that magnons typically repel near the plateau critical fields; however, near the critical value of $J'$ at which the classical plateau stabilizes, magnons with opposite spin and momentum attract, and bound-pair condensation preempts single-magnon condensation.  This condensation results in an exotic ``spin-current'' state that preserves both translational and $U(1)$ spin symmetry, but spontaneously breaks chiral symmetry.  We showed how to derive this state naturally via a relativistic scalar field theory with an O(4) field, and a consequent composite scalar Ising order parameter.

\begin{acknowledgments}
The authors would like to thank Oleg Starykh and Cenke Xu for helpful discussions, Hiroaki Ishizuka for translation from Japanese, and Sato Masahiro and Wei Zhu for bringing Refs.~\onlinecite{Sato11, Sato13, Zhu} to their attention.  They also thank the two anonymous reviewers for their helpful comments.  This work was supported by the NSF Materials Theory program under grant NSF DMR1506119.
\end{acknowledgments}

\bibliography{Biblio}

\begin{thebibliography}{53}%
\makeatletter
\providecommand \@ifxundefined [1]{%
 \@ifx{#1\undefined}
}%
\providecommand \@ifnum [1]{%
 \ifnum #1\expandafter \@firstoftwo
 \else \expandafter \@secondoftwo
 \fi
}%
\providecommand \@ifx [1]{%
 \ifx #1\expandafter \@firstoftwo
 \else \expandafter \@secondoftwo
 \fi
}%
\providecommand \natexlab [1]{#1}%
\providecommand \enquote  [1]{``#1''}%
\providecommand \bibnamefont  [1]{#1}%
\providecommand \bibfnamefont [1]{#1}%
\providecommand \citenamefont [1]{#1}%
\providecommand \href@noop [0]{\@secondoftwo}%
\providecommand \href [0]{\begingroup \@sanitize@url \@href}%
\providecommand \@href[1]{\@@startlink{#1}\@@href}%
\providecommand \@@href[1]{\endgroup#1\@@endlink}%
\providecommand \@sanitize@url [0]{\catcode `\\12\catcode `\$12\catcode
  `\&12\catcode `\#12\catcode `\^12\catcode `\_12\catcode `\%12\relax}%
\providecommand \@@startlink[1]{}%
\providecommand \@@endlink[0]{}%
\providecommand \url  [0]{\begingroup\@sanitize@url \@url }%
\providecommand \@url [1]{\endgroup\@href {#1}{\urlprefix }}%
\providecommand \urlprefix  [0]{URL }%
\providecommand \Eprint [0]{\href }%
\providecommand \doibase [0]{http://dx.doi.org/}%
\providecommand \selectlanguage [0]{\@gobble}%
\providecommand \bibinfo  [0]{\@secondoftwo}%
\providecommand \bibfield  [0]{\@secondoftwo}%
\providecommand \translation [1]{[#1]}%
\providecommand \BibitemOpen [0]{}%
\providecommand \bibitemStop [0]{}%
\providecommand \bibitemNoStop [0]{.\EOS\space}%
\providecommand \EOS [0]{\spacefactor3000\relax}%
\providecommand \BibitemShut  [1]{\csname bibitem#1\endcsname}%
\let\auto@bib@innerbib\@empty
\bibitem [{\citenamefont {Balents}(2010)}]{BalentsSL}%
  \BibitemOpen
  \bibfield  {author} {\bibinfo {author} {\bibfnamefont {L.}~\bibnamefont
  {Balents}},\ }\href@noop {} {\bibfield  {journal} {\bibinfo  {journal}
  {Nature}\ }\textbf {\bibinfo {volume} {464}},\ \bibinfo {pages} {199}
  (\bibinfo {year} {2010})}\BibitemShut {NoStop}%
\bibitem [{\citenamefont {Starykh}(2015)}]{Starykh}%
  \BibitemOpen
  \bibfield  {author} {\bibinfo {author} {\bibfnamefont {O.~A.}\ \bibnamefont
  {Starykh}},\ }\href@noop {} {\bibfield  {journal} {\bibinfo  {journal}
  {Reports on Progress in Physics}\ }\textbf {\bibinfo {volume} {78}} (\bibinfo
  {year} {2015})}\BibitemShut {NoStop}%
\bibitem [{\citenamefont {Rice}(2002)}]{Rice}%
  \BibitemOpen
  \bibfield  {author} {\bibinfo {author} {\bibfnamefont {T.~M.}\ \bibnamefont
  {Rice}},\ }\href {\doibase 10.1126/science.1078819} {\bibfield  {journal}
  {\bibinfo  {journal} {Science}\ }\textbf {\bibinfo {volume} {298}},\ \bibinfo
  {pages} {760} (\bibinfo {year} {2002})}\BibitemShut {NoStop}%
\bibitem [{\citenamefont {Giamarchi}\ \emph {et~al.}(2008)\citenamefont
  {Giamarchi}, \citenamefont {R{\"u}egg},\ and\ \citenamefont
  {Tchernyshyov}}]{Giamarchi}%
  \BibitemOpen
  \bibfield  {author} {\bibinfo {author} {\bibfnamefont {T.}~\bibnamefont
  {Giamarchi}}, \bibinfo {author} {\bibfnamefont {C.}~\bibnamefont
  {R{\"u}egg}}, \ and\ \bibinfo {author} {\bibfnamefont {O.}~\bibnamefont
  {Tchernyshyov}},\ }\href@noop {} {\bibfield  {journal} {\bibinfo  {journal}
  {Nature Physics}\ }\textbf {\bibinfo {volume} {4}},\ \bibinfo {pages} {198 }
  (\bibinfo {year} {2008})}\BibitemShut {NoStop}%
\bibitem [{\citenamefont {Alicea}\ \emph {et~al.}(2009)\citenamefont {Alicea},
  \citenamefont {Chubukov},\ and\ \citenamefont {Starykh}}]{Alicea}%
  \BibitemOpen
  \bibfield  {author} {\bibinfo {author} {\bibfnamefont {J.}~\bibnamefont
  {Alicea}}, \bibinfo {author} {\bibfnamefont {A.~V.}\ \bibnamefont
  {Chubukov}}, \ and\ \bibinfo {author} {\bibfnamefont {O.~A.}\ \bibnamefont
  {Starykh}},\ }\href {\doibase 10.1103/PhysRevLett.102.137201} {\bibfield
  {journal} {\bibinfo  {journal} {Phys. Rev. Lett.}\ }\textbf {\bibinfo
  {volume} {102}},\ \bibinfo {pages} {137201} (\bibinfo {year}
  {2009})}\BibitemShut {NoStop}%
\bibitem [{\citenamefont {Chubukov}\ and\ \citenamefont
  {Starykh}(2013)}]{ChubukovStarykh}%
  \BibitemOpen
  \bibfield  {author} {\bibinfo {author} {\bibfnamefont {A.~V.}\ \bibnamefont
  {Chubukov}}\ and\ \bibinfo {author} {\bibfnamefont {O.~A.}\ \bibnamefont
  {Starykh}},\ }\href {\doibase 10.1103/PhysRevLett.110.217210} {\bibfield
  {journal} {\bibinfo  {journal} {Phys. Rev. Lett.}\ }\textbf {\bibinfo
  {volume} {110}},\ \bibinfo {pages} {217210} (\bibinfo {year}
  {2013})}\BibitemShut {NoStop}%
\bibitem [{\citenamefont {Ono}\ \emph {et~al.}(2003)\citenamefont {Ono},
  \citenamefont {Tanaka}, \citenamefont {Aruga~Katori}, \citenamefont
  {Ishikawa}, \citenamefont {Mitamura},\ and\ \citenamefont {Goto}}]{Ono}%
  \BibitemOpen
  \bibfield  {author} {\bibinfo {author} {\bibfnamefont {T.}~\bibnamefont
  {Ono}}, \bibinfo {author} {\bibfnamefont {H.}~\bibnamefont {Tanaka}},
  \bibinfo {author} {\bibfnamefont {H.}~\bibnamefont {Aruga~Katori}}, \bibinfo
  {author} {\bibfnamefont {F.}~\bibnamefont {Ishikawa}}, \bibinfo {author}
  {\bibfnamefont {H.}~\bibnamefont {Mitamura}}, \ and\ \bibinfo {author}
  {\bibfnamefont {T.}~\bibnamefont {Goto}},\ }\href {\doibase
  10.1103/PhysRevB.67.104431} {\bibfield  {journal} {\bibinfo  {journal} {Phys.
  Rev. B}\ }\textbf {\bibinfo {volume} {67}},\ \bibinfo {pages} {104431}
  (\bibinfo {year} {2003})}\BibitemShut {NoStop}%
\bibitem [{\citenamefont {Tsujii}\ \emph {et~al.}(2007)\citenamefont {Tsujii},
  \citenamefont {Rotundu}, \citenamefont {Ono}, \citenamefont {Tanaka},
  \citenamefont {Andraka}, \citenamefont {Ingersent},\ and\ \citenamefont
  {Takano}}]{Tsujii}%
  \BibitemOpen
  \bibfield  {author} {\bibinfo {author} {\bibfnamefont {H.}~\bibnamefont
  {Tsujii}}, \bibinfo {author} {\bibfnamefont {C.~R.}\ \bibnamefont {Rotundu}},
  \bibinfo {author} {\bibfnamefont {T.}~\bibnamefont {Ono}}, \bibinfo {author}
  {\bibfnamefont {H.}~\bibnamefont {Tanaka}}, \bibinfo {author} {\bibfnamefont
  {B.}~\bibnamefont {Andraka}}, \bibinfo {author} {\bibfnamefont
  {K.}~\bibnamefont {Ingersent}}, \ and\ \bibinfo {author} {\bibfnamefont
  {Y.}~\bibnamefont {Takano}},\ }\href {\doibase 10.1103/PhysRevB.76.060406}
  {\bibfield  {journal} {\bibinfo  {journal} {Phys. Rev. B}\ }\textbf {\bibinfo
  {volume} {76}},\ \bibinfo {pages} {060406} (\bibinfo {year}
  {2007})}\BibitemShut {NoStop}%
\bibitem [{\citenamefont {Fortune}\ \emph {et~al.}(2009)\citenamefont
  {Fortune}, \citenamefont {Hannahs}, \citenamefont {Yoshida}, \citenamefont
  {Sherline}, \citenamefont {Ono}, \citenamefont {Tanaka},\ and\ \citenamefont
  {Takano}}]{Fortune}%
  \BibitemOpen
  \bibfield  {author} {\bibinfo {author} {\bibfnamefont {N.~A.}\ \bibnamefont
  {Fortune}}, \bibinfo {author} {\bibfnamefont {S.~T.}\ \bibnamefont
  {Hannahs}}, \bibinfo {author} {\bibfnamefont {Y.}~\bibnamefont {Yoshida}},
  \bibinfo {author} {\bibfnamefont {T.~E.}\ \bibnamefont {Sherline}}, \bibinfo
  {author} {\bibfnamefont {T.}~\bibnamefont {Ono}}, \bibinfo {author}
  {\bibfnamefont {H.}~\bibnamefont {Tanaka}}, \ and\ \bibinfo {author}
  {\bibfnamefont {Y.}~\bibnamefont {Takano}},\ }\href {\doibase
  10.1103/PhysRevLett.102.257201} {\bibfield  {journal} {\bibinfo  {journal}
  {Phys. Rev. Lett.}\ }\textbf {\bibinfo {volume} {102}},\ \bibinfo {pages}
  {257201} (\bibinfo {year} {2009})}\BibitemShut {NoStop}%
\bibitem [{\citenamefont {Ishikawa}\ \emph {et~al.}(2015)\citenamefont
  {Ishikawa}, \citenamefont {Yoshida}, \citenamefont {Nawa}, \citenamefont
  {Jeong}, \citenamefont {Kr\"amer}, \citenamefont
  {Horvati\ifmmode~\acute{c}\else \'{c}\fi{}}, \citenamefont {Berthier},
  \citenamefont {Takigawa}, \citenamefont {Akaki}, \citenamefont {Miyake},
  \citenamefont {Tokunaga}, \citenamefont {Kindo}, \citenamefont {Yamaura},
  \citenamefont {Okamoto},\ and\ \citenamefont {Hiroi}}]{Ishikawa}%
  \BibitemOpen
  \bibfield  {author} {\bibinfo {author} {\bibfnamefont {H.}~\bibnamefont
  {Ishikawa}}, \bibinfo {author} {\bibfnamefont {M.}~\bibnamefont {Yoshida}},
  \bibinfo {author} {\bibfnamefont {K.}~\bibnamefont {Nawa}}, \bibinfo {author}
  {\bibfnamefont {M.}~\bibnamefont {Jeong}}, \bibinfo {author} {\bibfnamefont
  {S.}~\bibnamefont {Kr\"amer}}, \bibinfo {author} {\bibfnamefont
  {M.}~\bibnamefont {Horvati\ifmmode~\acute{c}\else \'{c}\fi{}}}, \bibinfo
  {author} {\bibfnamefont {C.}~\bibnamefont {Berthier}}, \bibinfo {author}
  {\bibfnamefont {M.}~\bibnamefont {Takigawa}}, \bibinfo {author}
  {\bibfnamefont {M.}~\bibnamefont {Akaki}}, \bibinfo {author} {\bibfnamefont
  {A.}~\bibnamefont {Miyake}}, \bibinfo {author} {\bibfnamefont
  {M.}~\bibnamefont {Tokunaga}}, \bibinfo {author} {\bibfnamefont
  {K.}~\bibnamefont {Kindo}}, \bibinfo {author} {\bibfnamefont
  {J.}~\bibnamefont {Yamaura}}, \bibinfo {author} {\bibfnamefont
  {Y.}~\bibnamefont {Okamoto}}, \ and\ \bibinfo {author} {\bibfnamefont
  {Z.}~\bibnamefont {Hiroi}},\ }\href {\doibase 10.1103/PhysRevLett.114.227202}
  {\bibfield  {journal} {\bibinfo  {journal} {Phys. Rev. Lett.}\ }\textbf
  {\bibinfo {volume} {114}},\ \bibinfo {pages} {227202} (\bibinfo {year}
  {2015})}\BibitemShut {NoStop}%
\bibitem [{\citenamefont {Yamashita}\ \emph {et~al.}(2014)\citenamefont
  {Yamashita}, \citenamefont {Miyata}, \citenamefont {Takeyama}, \citenamefont
  {Ishikawa}, \citenamefont {Okamoto},\ and\ \citenamefont
  {Hiroi}}]{Yamashita}%
  \BibitemOpen
  \bibfield  {author} {\bibinfo {author} {\bibfnamefont {T.}~\bibnamefont
  {Yamashita}}, \bibinfo {author} {\bibfnamefont {A.}~\bibnamefont {Miyata}},
  \bibinfo {author} {\bibfnamefont {S.}~\bibnamefont {Takeyama}}, \bibinfo
  {author} {\bibfnamefont {H.}~\bibnamefont {Ishikawa}}, \bibinfo {author}
  {\bibfnamefont {Y.}~\bibnamefont {Okamoto}}, \ and\ \bibinfo {author}
  {\bibfnamefont {Z.}~\bibnamefont {Hiroi}},\ }\href
  {http://ci.nii.ac.jp/naid/110009835239/en/} {\bibfield  {journal} {\bibinfo
  {journal} {Meeting abstracts of the Physical Society of Japan}\ }\textbf
  {\bibinfo {volume} {69}},\ \bibinfo {pages} {471} (\bibinfo {year}
  {2014})}\BibitemShut {NoStop}%
\bibitem [{\citenamefont {Janson}\ \emph {et~al.}(2010)\citenamefont {Janson},
  \citenamefont {Richter}, \citenamefont {Sindzingre},\ and\ \citenamefont
  {Rosner}}]{Janson10}%
  \BibitemOpen
  \bibfield  {author} {\bibinfo {author} {\bibfnamefont {O.}~\bibnamefont
  {Janson}}, \bibinfo {author} {\bibfnamefont {J.}~\bibnamefont {Richter}},
  \bibinfo {author} {\bibfnamefont {P.}~\bibnamefont {Sindzingre}}, \ and\
  \bibinfo {author} {\bibfnamefont {H.}~\bibnamefont {Rosner}},\ }\href
  {\doibase 10.1103/PhysRevB.82.104434} {\bibfield  {journal} {\bibinfo
  {journal} {Phys. Rev. B}\ }\textbf {\bibinfo {volume} {82}},\ \bibinfo
  {pages} {104434} (\bibinfo {year} {2010})}\BibitemShut {NoStop}%
\bibitem [{\citenamefont {Janson}\ \emph {et~al.}(2016)\citenamefont {Janson},
  \citenamefont {Furukawa}, \citenamefont {Momoi}, \citenamefont {Sindzingre},
  \citenamefont {Richter},\ and\ \citenamefont {Held}}]{Janson16}%
  \BibitemOpen
  \bibfield  {author} {\bibinfo {author} {\bibfnamefont {O.}~\bibnamefont
  {Janson}}, \bibinfo {author} {\bibfnamefont {S.}~\bibnamefont {Furukawa}},
  \bibinfo {author} {\bibfnamefont {T.}~\bibnamefont {Momoi}}, \bibinfo
  {author} {\bibfnamefont {P.}~\bibnamefont {Sindzingre}}, \bibinfo {author}
  {\bibfnamefont {J.}~\bibnamefont {Richter}}, \ and\ \bibinfo {author}
  {\bibfnamefont {K.}~\bibnamefont {Held}},\ }\href {\doibase
  10.1103/PhysRevLett.117.037206} {\bibfield  {journal} {\bibinfo  {journal}
  {Phys. Rev. Lett.}\ }\textbf {\bibinfo {volume} {117}},\ \bibinfo {pages}
  {037206} (\bibinfo {year} {2016})}\BibitemShut {NoStop}%
\bibitem [{\citenamefont {Schulenburg}\ \emph {et~al.}(2002)\citenamefont
  {Schulenburg}, \citenamefont {Honecker}, \citenamefont {Schnack},
  \citenamefont {Richter},\ and\ \citenamefont {Schmidt}}]{Schulenburg}%
  \BibitemOpen
  \bibfield  {author} {\bibinfo {author} {\bibfnamefont {J.}~\bibnamefont
  {Schulenburg}}, \bibinfo {author} {\bibfnamefont {A.}~\bibnamefont
  {Honecker}}, \bibinfo {author} {\bibfnamefont {J.}~\bibnamefont {Schnack}},
  \bibinfo {author} {\bibfnamefont {J.}~\bibnamefont {Richter}}, \ and\
  \bibinfo {author} {\bibfnamefont {H.-J.}\ \bibnamefont {Schmidt}},\ }\href
  {\doibase 10.1103/PhysRevLett.88.167207} {\bibfield  {journal} {\bibinfo
  {journal} {Phys. Rev. Lett.}\ }\textbf {\bibinfo {volume} {88}},\ \bibinfo
  {pages} {167207} (\bibinfo {year} {2002})}\BibitemShut {NoStop}%
\bibitem [{\citenamefont {Nishimoto}\ \emph {et~al.}(2013)\citenamefont
  {Nishimoto}, \citenamefont {Shibata},\ and\ \citenamefont
  {Hotta}}]{Nishimoto}%
  \BibitemOpen
  \bibfield  {author} {\bibinfo {author} {\bibfnamefont {S.}~\bibnamefont
  {Nishimoto}}, \bibinfo {author} {\bibfnamefont {N.}~\bibnamefont {Shibata}},
  \ and\ \bibinfo {author} {\bibfnamefont {C.}~\bibnamefont {Hotta}},\
  }\href@noop {} {\bibfield  {journal} {\bibinfo  {journal} {Nature
  Communications}\ }\textbf {\bibinfo {volume} {4}} (\bibinfo {year}
  {2013})}\BibitemShut {NoStop}%
\bibitem [{\citenamefont {Capponi}\ \emph {et~al.}(2013)\citenamefont
  {Capponi}, \citenamefont {Derzhko}, \citenamefont {Honecker}, \citenamefont
  {L\"auchli},\ and\ \citenamefont {Richter}}]{Capponi}%
  \BibitemOpen
  \bibfield  {author} {\bibinfo {author} {\bibfnamefont {S.}~\bibnamefont
  {Capponi}}, \bibinfo {author} {\bibfnamefont {O.}~\bibnamefont {Derzhko}},
  \bibinfo {author} {\bibfnamefont {A.}~\bibnamefont {Honecker}}, \bibinfo
  {author} {\bibfnamefont {A.~M.}\ \bibnamefont {L\"auchli}}, \ and\ \bibinfo
  {author} {\bibfnamefont {J.}~\bibnamefont {Richter}},\ }\href {\doibase
  10.1103/PhysRevB.88.144416} {\bibfield  {journal} {\bibinfo  {journal} {Phys.
  Rev. B}\ }\textbf {\bibinfo {volume} {88}},\ \bibinfo {pages} {144416}
  (\bibinfo {year} {2013})}\BibitemShut {NoStop}%
\bibitem [{\citenamefont {Damle}\ and\ \citenamefont {Senthil}(2006)}]{Damle}%
  \BibitemOpen
  \bibfield  {author} {\bibinfo {author} {\bibfnamefont {K.}~\bibnamefont
  {Damle}}\ and\ \bibinfo {author} {\bibfnamefont {T.}~\bibnamefont
  {Senthil}},\ }\href {\doibase 10.1103/PhysRevLett.97.067202} {\bibfield
  {journal} {\bibinfo  {journal} {Phys. Rev. Lett.}\ }\textbf {\bibinfo
  {volume} {97}},\ \bibinfo {pages} {067202} (\bibinfo {year}
  {2006})}\BibitemShut {NoStop}%
\bibitem [{\citenamefont {Zhu}\ \emph {et~al.}(2016)\citenamefont {Zhu},
  \citenamefont {Gong},\ and\ \citenamefont {Sheng}}]{Zhu}%
  \BibitemOpen
  \bibfield  {author} {\bibinfo {author} {\bibfnamefont {W.}~\bibnamefont
  {Zhu}}, \bibinfo {author} {\bibfnamefont {S.~S.}\ \bibnamefont {Gong}}, \
  and\ \bibinfo {author} {\bibfnamefont {D.~N.}\ \bibnamefont {Sheng}},\ }\href
  {\doibase 10.1103/PhysRevB.94.035129} {\bibfield  {journal} {\bibinfo
  {journal} {Phys. Rev. B}\ }\textbf {\bibinfo {volume} {94}},\ \bibinfo
  {pages} {035129} (\bibinfo {year} {2016})}\BibitemShut {NoStop}%
\bibitem [{\citenamefont {Fisher}\ \emph {et~al.}(1989)\citenamefont {Fisher},
  \citenamefont {Weichman}, \citenamefont {Grinstein},\ and\ \citenamefont
  {Fisher}}]{Fisher}%
  \BibitemOpen
  \bibfield  {author} {\bibinfo {author} {\bibfnamefont {M.~P.~A.}\
  \bibnamefont {Fisher}}, \bibinfo {author} {\bibfnamefont {P.~B.}\
  \bibnamefont {Weichman}}, \bibinfo {author} {\bibfnamefont {G.}~\bibnamefont
  {Grinstein}}, \ and\ \bibinfo {author} {\bibfnamefont {D.~S.}\ \bibnamefont
  {Fisher}},\ }\href {\doibase 10.1103/PhysRevB.40.546} {\bibfield  {journal}
  {\bibinfo  {journal} {Phys. Rev. B}\ }\textbf {\bibinfo {volume} {40}},\
  \bibinfo {pages} {546} (\bibinfo {year} {1989})}\BibitemShut {NoStop}%
\bibitem [{\citenamefont {Chubukov}(1991)}]{Chubukov}%
  \BibitemOpen
  \bibfield  {author} {\bibinfo {author} {\bibfnamefont {A.~V.}\ \bibnamefont
  {Chubukov}},\ }\href {\doibase 10.1103/PhysRevB.44.4693} {\bibfield
  {journal} {\bibinfo  {journal} {Phys. Rev. B}\ }\textbf {\bibinfo {volume}
  {44}},\ \bibinfo {pages} {4693} (\bibinfo {year} {1991})}\BibitemShut
  {NoStop}%
\bibitem [{\citenamefont {Heidrich-Meisner}\ \emph {et~al.}(2006)\citenamefont
  {Heidrich-Meisner}, \citenamefont {Honecker},\ and\ \citenamefont
  {Vekua}}]{Heidrich-Meisner}%
  \BibitemOpen
  \bibfield  {author} {\bibinfo {author} {\bibfnamefont {F.}~\bibnamefont
  {Heidrich-Meisner}}, \bibinfo {author} {\bibfnamefont {A.}~\bibnamefont
  {Honecker}}, \ and\ \bibinfo {author} {\bibfnamefont {T.}~\bibnamefont
  {Vekua}},\ }\href {\doibase 10.1103/PhysRevB.74.020403} {\bibfield  {journal}
  {\bibinfo  {journal} {Phys. Rev. B}\ }\textbf {\bibinfo {volume} {74}},\
  \bibinfo {pages} {020403} (\bibinfo {year} {2006})}\BibitemShut {NoStop}%
\bibitem [{\citenamefont {Kuzian}\ and\ \citenamefont
  {Drechsler}(2007)}]{Kuzian}%
  \BibitemOpen
  \bibfield  {author} {\bibinfo {author} {\bibfnamefont {R.~O.}\ \bibnamefont
  {Kuzian}}\ and\ \bibinfo {author} {\bibfnamefont {S.-L.}\ \bibnamefont
  {Drechsler}},\ }\href {\doibase 10.1103/PhysRevB.75.024401} {\bibfield
  {journal} {\bibinfo  {journal} {Phys. Rev. B}\ }\textbf {\bibinfo {volume}
  {75}},\ \bibinfo {pages} {024401} (\bibinfo {year} {2007})}\BibitemShut
  {NoStop}%
\bibitem [{\citenamefont {Kecke}\ \emph {et~al.}(2007)\citenamefont {Kecke},
  \citenamefont {Momoi},\ and\ \citenamefont {Furusaki}}]{Kecke}%
  \BibitemOpen
  \bibfield  {author} {\bibinfo {author} {\bibfnamefont {L.}~\bibnamefont
  {Kecke}}, \bibinfo {author} {\bibfnamefont {T.}~\bibnamefont {Momoi}}, \ and\
  \bibinfo {author} {\bibfnamefont {A.}~\bibnamefont {Furusaki}},\ }\href
  {\doibase 10.1103/PhysRevB.76.060407} {\bibfield  {journal} {\bibinfo
  {journal} {Phys. Rev. B}\ }\textbf {\bibinfo {volume} {76}},\ \bibinfo
  {pages} {060407} (\bibinfo {year} {2007})}\BibitemShut {NoStop}%
\bibitem [{\citenamefont {Hikihara}\ \emph {et~al.}(2008)\citenamefont
  {Hikihara}, \citenamefont {Kecke}, \citenamefont {Momoi},\ and\ \citenamefont
  {Furusaki}}]{Hikihara}%
  \BibitemOpen
  \bibfield  {author} {\bibinfo {author} {\bibfnamefont {T.}~\bibnamefont
  {Hikihara}}, \bibinfo {author} {\bibfnamefont {L.}~\bibnamefont {Kecke}},
  \bibinfo {author} {\bibfnamefont {T.}~\bibnamefont {Momoi}}, \ and\ \bibinfo
  {author} {\bibfnamefont {A.}~\bibnamefont {Furusaki}},\ }\href {\doibase
  10.1103/PhysRevB.78.144404} {\bibfield  {journal} {\bibinfo  {journal} {Phys.
  Rev. B}\ }\textbf {\bibinfo {volume} {78}},\ \bibinfo {pages} {144404}
  (\bibinfo {year} {2008})}\BibitemShut {NoStop}%
\bibitem [{\citenamefont {Sudan}\ \emph {et~al.}(2009)\citenamefont {Sudan},
  \citenamefont {L\"uscher},\ and\ \citenamefont {L\"auchli}}]{Sudan}%
  \BibitemOpen
  \bibfield  {author} {\bibinfo {author} {\bibfnamefont {J.}~\bibnamefont
  {Sudan}}, \bibinfo {author} {\bibfnamefont {A.}~\bibnamefont {L\"uscher}}, \
  and\ \bibinfo {author} {\bibfnamefont {A.~M.}\ \bibnamefont {L\"auchli}},\
  }\href {\doibase 10.1103/PhysRevB.80.140402} {\bibfield  {journal} {\bibinfo
  {journal} {Phys. Rev. B}\ }\textbf {\bibinfo {volume} {80}},\ \bibinfo
  {pages} {140402} (\bibinfo {year} {2009})}\BibitemShut {NoStop}%
\bibitem [{\citenamefont {Sato}\ \emph {et~al.}(2011)\citenamefont {Sato},
  \citenamefont {Hikihara},\ and\ \citenamefont {Momoi}}]{Sato11}%
  \BibitemOpen
  \bibfield  {author} {\bibinfo {author} {\bibfnamefont {M.}~\bibnamefont
  {Sato}}, \bibinfo {author} {\bibfnamefont {T.}~\bibnamefont {Hikihara}}, \
  and\ \bibinfo {author} {\bibfnamefont {T.}~\bibnamefont {Momoi}},\ }\href
  {\doibase 10.1103/PhysRevB.83.064405} {\bibfield  {journal} {\bibinfo
  {journal} {Phys. Rev. B}\ }\textbf {\bibinfo {volume} {83}},\ \bibinfo
  {pages} {064405} (\bibinfo {year} {2011})}\BibitemShut {NoStop}%
\bibitem [{\citenamefont {Sato}\ \emph {et~al.}(2013)\citenamefont {Sato},
  \citenamefont {Hikihara},\ and\ \citenamefont {Momoi}}]{Sato13}%
  \BibitemOpen
  \bibfield  {author} {\bibinfo {author} {\bibfnamefont {M.}~\bibnamefont
  {Sato}}, \bibinfo {author} {\bibfnamefont {T.}~\bibnamefont {Hikihara}}, \
  and\ \bibinfo {author} {\bibfnamefont {T.}~\bibnamefont {Momoi}},\ }\href
  {\doibase 10.1103/PhysRevLett.110.077206} {\bibfield  {journal} {\bibinfo
  {journal} {Phys. Rev. Lett.}\ }\textbf {\bibinfo {volume} {110}},\ \bibinfo
  {pages} {077206} (\bibinfo {year} {2013})}\BibitemShut {NoStop}%
\bibitem [{\citenamefont {Balents}\ and\ \citenamefont
  {Starykh}(2016)}]{Balents}%
  \BibitemOpen
  \bibfield  {author} {\bibinfo {author} {\bibfnamefont {L.}~\bibnamefont
  {Balents}}\ and\ \bibinfo {author} {\bibfnamefont {O.~A.}\ \bibnamefont
  {Starykh}},\ }\href {\doibase 10.1103/PhysRevLett.116.177201} {\bibfield
  {journal} {\bibinfo  {journal} {Phys. Rev. Lett.}\ }\textbf {\bibinfo
  {volume} {116}},\ \bibinfo {pages} {177201} (\bibinfo {year}
  {2016})}\BibitemShut {NoStop}%
\bibitem [{\citenamefont {Luttinger}\ and\ \citenamefont
  {Tisza}(1946)}]{Luttinger}%
  \BibitemOpen
  \bibfield  {author} {\bibinfo {author} {\bibfnamefont {J.~M.}\ \bibnamefont
  {Luttinger}}\ and\ \bibinfo {author} {\bibfnamefont {L.}~\bibnamefont
  {Tisza}},\ }\href {\doibase 10.1103/PhysRev.70.954} {\bibfield  {journal}
  {\bibinfo  {journal} {Phys. Rev.}\ }\textbf {\bibinfo {volume} {70}},\
  \bibinfo {pages} {954} (\bibinfo {year} {1946})}\BibitemShut {NoStop}%
\bibitem [{\citenamefont {Kaplan}\ and\ \citenamefont {Menyuk}(2007)}]{Kaplan}%
  \BibitemOpen
  \bibfield  {author} {\bibinfo {author} {\bibfnamefont {T.~A.}\ \bibnamefont
  {Kaplan}}\ and\ \bibinfo {author} {\bibfnamefont {N.}~\bibnamefont
  {Menyuk}},\ }\href@noop {} {\bibfield  {journal} {\bibinfo  {journal}
  {Philosophical Magazine}\ }\textbf {\bibinfo {volume} {87}},\ \bibinfo
  {pages} {3711} (\bibinfo {year} {2007})}\BibitemShut {NoStop}%
\bibitem [{\citenamefont {Litvin}(1974)}]{Litvin}%
  \BibitemOpen
  \bibfield  {author} {\bibinfo {author} {\bibfnamefont {D.~B.}\ \bibnamefont
  {Litvin}},\ }\href@noop {} {\bibfield  {journal} {\bibinfo  {journal}
  {Physica}\ }\textbf {\bibinfo {volume} {77}},\ \bibinfo {pages} {205}
  (\bibinfo {year} {1974})}\BibitemShut {NoStop}%
\bibitem [{\citenamefont {Xiong}\ and\ \citenamefont {Wen}()}]{Xiong}%
  \BibitemOpen
  \bibfield  {author} {\bibinfo {author} {\bibfnamefont {Z.}~\bibnamefont
  {Xiong}}\ and\ \bibinfo {author} {\bibfnamefont {X.-G.}\ \bibnamefont
  {Wen}},\ }\href@noop {} {}\bibinfo {note} {1208.1512
  [cond-mat.stat-mech]}\BibitemShut {NoStop}%
\bibitem [{app()}]{app}%
  \BibitemOpen
  \href@noop {} {}\bibinfo {note} {See appendices.}\BibitemShut {Stop}%
\bibitem [{\citenamefont {Holstein}\ and\ \citenamefont
  {Primakoff}(1940)}]{HP}%
  \BibitemOpen
  \bibfield  {author} {\bibinfo {author} {\bibfnamefont {T.}~\bibnamefont
  {Holstein}}\ and\ \bibinfo {author} {\bibfnamefont {H.}~\bibnamefont
  {Primakoff}},\ }\href {\doibase 10.1103/PhysRev.58.1098} {\bibfield
  {journal} {\bibinfo  {journal} {Phys. Rev.}\ }\textbf {\bibinfo {volume}
  {58}},\ \bibinfo {pages} {1098} (\bibinfo {year} {1940})}\BibitemShut
  {NoStop}%
\bibitem [{\citenamefont {Oguchi}(1960)}]{Oguchi}%
  \BibitemOpen
  \bibfield  {author} {\bibinfo {author} {\bibfnamefont {T.}~\bibnamefont
  {Oguchi}},\ }\href {\doibase 10.1103/PhysRev.117.117} {\bibfield  {journal}
  {\bibinfo  {journal} {Phys. Rev.}\ }\textbf {\bibinfo {volume} {117}},\
  \bibinfo {pages} {117} (\bibinfo {year} {1960})}\BibitemShut {NoStop}%
\bibitem [{\citenamefont {Henley}(1989)}]{Henley}%
  \BibitemOpen
  \bibfield  {author} {\bibinfo {author} {\bibfnamefont {C.~L.}\ \bibnamefont
  {Henley}},\ }\href {\doibase 10.1103/PhysRevLett.62.2056} {\bibfield
  {journal} {\bibinfo  {journal} {Phys. Rev. Lett.}\ }\textbf {\bibinfo
  {volume} {62}},\ \bibinfo {pages} {2056} (\bibinfo {year}
  {1989})}\BibitemShut {NoStop}%
\bibitem [{\citenamefont {Liu}\ and\ \citenamefont {Fisher}(1973)}]{Liu}%
  \BibitemOpen
  \bibfield  {author} {\bibinfo {author} {\bibfnamefont {K.-S.}\ \bibnamefont
  {Liu}}\ and\ \bibinfo {author} {\bibfnamefont {M.~E.}\ \bibnamefont
  {Fisher}},\ }\href {\doibase 10.1007/BF00655458} {\bibfield  {journal}
  {\bibinfo  {journal} {Journal of Low Temperature Physics}\ }\textbf {\bibinfo
  {volume} {10}},\ \bibinfo {pages} {655} (\bibinfo {year} {1973})}\BibitemShut
  {NoStop}%
\bibitem [{\citenamefont {Kosterlitz}\ \emph {et~al.}(1976)\citenamefont
  {Kosterlitz}, \citenamefont {Nelson},\ and\ \citenamefont
  {Fisher}}]{Kosterlitz}%
  \BibitemOpen
  \bibfield  {author} {\bibinfo {author} {\bibfnamefont {J.~M.}\ \bibnamefont
  {Kosterlitz}}, \bibinfo {author} {\bibfnamefont {D.~R.}\ \bibnamefont
  {Nelson}}, \ and\ \bibinfo {author} {\bibfnamefont {M.~E.}\ \bibnamefont
  {Fisher}},\ }\href {\doibase 10.1103/PhysRevB.13.412} {\bibfield  {journal}
  {\bibinfo  {journal} {Phys. Rev. B}\ }\textbf {\bibinfo {volume} {13}},\
  \bibinfo {pages} {412} (\bibinfo {year} {1976})}\BibitemShut {NoStop}%
\bibitem [{\citenamefont {Calabrese}\ \emph {et~al.}(2003)\citenamefont
  {Calabrese}, \citenamefont {Pelissetto},\ and\ \citenamefont
  {Vicari}}]{Calabrese03}%
  \BibitemOpen
  \bibfield  {author} {\bibinfo {author} {\bibfnamefont {P.}~\bibnamefont
  {Calabrese}}, \bibinfo {author} {\bibfnamefont {A.}~\bibnamefont
  {Pelissetto}}, \ and\ \bibinfo {author} {\bibfnamefont {E.}~\bibnamefont
  {Vicari}},\ }\href {\doibase 10.1103/PhysRevB.67.054505} {\bibfield
  {journal} {\bibinfo  {journal} {Phys. Rev. B}\ }\textbf {\bibinfo {volume}
  {67}},\ \bibinfo {pages} {054505} (\bibinfo {year} {2003})}\BibitemShut
  {NoStop}%
\bibitem [{\citenamefont {Calabrese}\ \emph {et~al.}(2002)\citenamefont
  {Calabrese}, \citenamefont {Pelissetto},\ and\ \citenamefont
  {Vicari}}]{Calabrese02}%
  \BibitemOpen
  \bibfield  {author} {\bibinfo {author} {\bibfnamefont {P.}~\bibnamefont
  {Calabrese}}, \bibinfo {author} {\bibfnamefont {A.}~\bibnamefont
  {Pelissetto}}, \ and\ \bibinfo {author} {\bibfnamefont {E.}~\bibnamefont
  {Vicari}},\ }\href {\doibase 10.1103/PhysRevE.65.046115} {\bibfield
  {journal} {\bibinfo  {journal} {Phys. Rev. E}\ }\textbf {\bibinfo {volume}
  {65}},\ \bibinfo {pages} {046115} (\bibinfo {year} {2002})}\BibitemShut
  {NoStop}%
\bibitem [{\citenamefont {Kennedy}(1991)}]{Kennedy}%
  \BibitemOpen
  \bibfield  {author} {\bibinfo {author} {\bibfnamefont {T.}~\bibnamefont
  {Kennedy}},\ }\href {\doibase 10.1007/BF02100280} {\bibfield  {journal}
  {\bibinfo  {journal} {Communications in Mathematical Physics}\ }\textbf
  {\bibinfo {volume} {137}},\ \bibinfo {pages} {599} (\bibinfo {year}
  {1991})}\BibitemShut {NoStop}%
\bibitem [{\citenamefont {Rams}\ \emph {et~al.}(2015)\citenamefont {Rams},
  \citenamefont {Zauner}, \citenamefont {Bal}, \citenamefont {Haegeman},\ and\
  \citenamefont {Verstraete}}]{Rams}%
  \BibitemOpen
  \bibfield  {author} {\bibinfo {author} {\bibfnamefont {M.~M.}\ \bibnamefont
  {Rams}}, \bibinfo {author} {\bibfnamefont {V.}~\bibnamefont {Zauner}},
  \bibinfo {author} {\bibfnamefont {M.}~\bibnamefont {Bal}}, \bibinfo {author}
  {\bibfnamefont {J.}~\bibnamefont {Haegeman}}, \ and\ \bibinfo {author}
  {\bibfnamefont {F.}~\bibnamefont {Verstraete}},\ }\href {\doibase
  10.1103/PhysRevB.92.235150} {\bibfield  {journal} {\bibinfo  {journal} {Phys.
  Rev. B}\ }\textbf {\bibinfo {volume} {92}},\ \bibinfo {pages} {235150}
  (\bibinfo {year} {2015})}\BibitemShut {NoStop}%
\bibitem [{\citenamefont {Mukamel}\ and\ \citenamefont
  {Krinsky}(1976)}]{Mukamel}%
  \BibitemOpen
  \bibfield  {author} {\bibinfo {author} {\bibfnamefont {D.}~\bibnamefont
  {Mukamel}}\ and\ \bibinfo {author} {\bibfnamefont {S.}~\bibnamefont
  {Krinsky}},\ }\href {\doibase 10.1103/PhysRevB.13.5078} {\bibfield  {journal}
  {\bibinfo  {journal} {Phys. Rev. B}\ }\textbf {\bibinfo {volume} {13}},\
  \bibinfo {pages} {5078} (\bibinfo {year} {1976})}\BibitemShut {NoStop}%
\bibitem [{\citenamefont {Domany}\ \emph {et~al.}(1977)\citenamefont {Domany},
  \citenamefont {Mukamel},\ and\ \citenamefont {Fisher}}]{Domany}%
  \BibitemOpen
  \bibfield  {author} {\bibinfo {author} {\bibfnamefont {E.}~\bibnamefont
  {Domany}}, \bibinfo {author} {\bibfnamefont {D.}~\bibnamefont {Mukamel}}, \
  and\ \bibinfo {author} {\bibfnamefont {M.~E.}\ \bibnamefont {Fisher}},\
  }\href {\doibase 10.1103/PhysRevB.15.5432} {\bibfield  {journal} {\bibinfo
  {journal} {Phys. Rev. B}\ }\textbf {\bibinfo {volume} {15}},\ \bibinfo
  {pages} {5432} (\bibinfo {year} {1977})}\BibitemShut {NoStop}%
\bibitem [{\citenamefont {Cheong}\ and\ \citenamefont
  {Mostovoy}(2007)}]{Cheong}%
  \BibitemOpen
  \bibfield  {author} {\bibinfo {author} {\bibfnamefont {S.-W.}\ \bibnamefont
  {Cheong}}\ and\ \bibinfo {author} {\bibfnamefont {M.}~\bibnamefont
  {Mostovoy}},\ }\href@noop {} {\bibfield  {journal} {\bibinfo  {journal}
  {Nature Materials}\ }\textbf {\bibinfo {volume} {6}},\ \bibinfo {pages} {13}
  (\bibinfo {year} {2007})}\BibitemShut {NoStop}%
\bibitem [{\citenamefont {Dzyaloshinsky}(1958)}]{Dzyaloshinskii}%
  \BibitemOpen
  \bibfield  {author} {\bibinfo {author} {\bibfnamefont {I.}~\bibnamefont
  {Dzyaloshinsky}},\ }\href@noop {} {\bibfield  {journal} {\bibinfo  {journal}
  {Journal of Physics and Chemistry of Solids}\ }\textbf {\bibinfo {volume}
  {4}},\ \bibinfo {pages} {241Ð255} (\bibinfo {year} {1958})}\BibitemShut
  {NoStop}%
\bibitem [{\citenamefont {Moriya}(1960)}]{Moriya}%
  \BibitemOpen
  \bibfield  {author} {\bibinfo {author} {\bibfnamefont {T.}~\bibnamefont
  {Moriya}},\ }\href {\doibase 10.1103/PhysRev.120.91} {\bibfield  {journal}
  {\bibinfo  {journal} {Phys. Rev.}\ }\textbf {\bibinfo {volume} {120}},\
  \bibinfo {pages} {91} (\bibinfo {year} {1960})}\BibitemShut {NoStop}%
\bibitem [{\citenamefont {Katsura}\ \emph {et~al.}(2005)\citenamefont
  {Katsura}, \citenamefont {Nagaosa},\ and\ \citenamefont
  {Balatsky}}]{Katsura}%
  \BibitemOpen
  \bibfield  {author} {\bibinfo {author} {\bibfnamefont {H.}~\bibnamefont
  {Katsura}}, \bibinfo {author} {\bibfnamefont {N.}~\bibnamefont {Nagaosa}}, \
  and\ \bibinfo {author} {\bibfnamefont {A.~V.}\ \bibnamefont {Balatsky}},\
  }\href {\doibase 10.1103/PhysRevLett.95.057205} {\bibfield  {journal}
  {\bibinfo  {journal} {Phys. Rev. Lett.}\ }\textbf {\bibinfo {volume} {95}},\
  \bibinfo {pages} {057205} (\bibinfo {year} {2005})}\BibitemShut {NoStop}%
\bibitem [{\citenamefont {Mostovoy}(2006)}]{Mostovoy}%
  \BibitemOpen
  \bibfield  {author} {\bibinfo {author} {\bibfnamefont {M.}~\bibnamefont
  {Mostovoy}},\ }\href {\doibase 10.1103/PhysRevLett.96.067601} {\bibfield
  {journal} {\bibinfo  {journal} {Phys. Rev. Lett.}\ }\textbf {\bibinfo
  {volume} {96}},\ \bibinfo {pages} {067601} (\bibinfo {year}
  {2006})}\BibitemShut {NoStop}%
\bibitem [{\citenamefont {Al-Hassanieh}\ \emph {et~al.}(2009)\citenamefont
  {Al-Hassanieh}, \citenamefont {Batista}, \citenamefont {Ortiz},\ and\
  \citenamefont {Bulaevskii}}]{Al-Hassanieh}%
  \BibitemOpen
  \bibfield  {author} {\bibinfo {author} {\bibfnamefont {K.~A.}\ \bibnamefont
  {Al-Hassanieh}}, \bibinfo {author} {\bibfnamefont {C.~D.}\ \bibnamefont
  {Batista}}, \bibinfo {author} {\bibfnamefont {G.}~\bibnamefont {Ortiz}}, \
  and\ \bibinfo {author} {\bibfnamefont {L.~N.}\ \bibnamefont {Bulaevskii}},\
  }\href {\doibase 10.1103/PhysRevLett.103.216402} {\bibfield  {journal}
  {\bibinfo  {journal} {Phys. Rev. Lett.}\ }\textbf {\bibinfo {volume} {103}},\
  \bibinfo {pages} {216402} (\bibinfo {year} {2009})}\BibitemShut {NoStop}%
\bibitem [{\citenamefont {Onose}\ \emph {et~al.}(2010)\citenamefont {Onose},
  \citenamefont {Ideue}, \citenamefont {Katsura}, \citenamefont {Shiomi},
  \citenamefont {Nagaosa},\ and\ \citenamefont {Tokura}}]{Onose}%
  \BibitemOpen
  \bibfield  {author} {\bibinfo {author} {\bibfnamefont {Y.}~\bibnamefont
  {Onose}}, \bibinfo {author} {\bibfnamefont {T.}~\bibnamefont {Ideue}},
  \bibinfo {author} {\bibfnamefont {H.}~\bibnamefont {Katsura}}, \bibinfo
  {author} {\bibfnamefont {Y.}~\bibnamefont {Shiomi}}, \bibinfo {author}
  {\bibfnamefont {N.}~\bibnamefont {Nagaosa}}, \ and\ \bibinfo {author}
  {\bibfnamefont {Y.}~\bibnamefont {Tokura}},\ }\href {\doibase
  10.1126/science.1188260} {\bibfield  {journal} {\bibinfo  {journal}
  {Science}\ }\textbf {\bibinfo {volume} {329}},\ \bibinfo {pages} {297}
  (\bibinfo {year} {2010})}\BibitemShut {NoStop}%
\bibitem [{\citenamefont {Arlego}\ \emph {et~al.}(2011)\citenamefont {Arlego},
  \citenamefont {Heidrich-Meisner}, \citenamefont {Honecker}, \citenamefont
  {Rossini},\ and\ \citenamefont {Vekua}}]{Arlego}%
  \BibitemOpen
  \bibfield  {author} {\bibinfo {author} {\bibfnamefont {M.}~\bibnamefont
  {Arlego}}, \bibinfo {author} {\bibfnamefont {F.}~\bibnamefont
  {Heidrich-Meisner}}, \bibinfo {author} {\bibfnamefont {A.}~\bibnamefont
  {Honecker}}, \bibinfo {author} {\bibfnamefont {G.}~\bibnamefont {Rossini}}, \
  and\ \bibinfo {author} {\bibfnamefont {T.}~\bibnamefont {Vekua}},\ }\href
  {\doibase 10.1103/PhysRevB.84.224409} {\bibfield  {journal} {\bibinfo
  {journal} {Phys. Rev. B}\ }\textbf {\bibinfo {volume} {84}},\ \bibinfo
  {pages} {224409} (\bibinfo {year} {2011})}\BibitemShut {NoStop}%
\bibitem [{\citenamefont {Kolezhuk}\ \emph {et~al.}(2012)\citenamefont
  {Kolezhuk}, \citenamefont {Heidrich-Meisner}, \citenamefont {Greschner},\
  and\ \citenamefont {Vekua}}]{Kolezhuk}%
  \BibitemOpen
  \bibfield  {author} {\bibinfo {author} {\bibfnamefont {A.~K.}\ \bibnamefont
  {Kolezhuk}}, \bibinfo {author} {\bibfnamefont {F.}~\bibnamefont
  {Heidrich-Meisner}}, \bibinfo {author} {\bibfnamefont {S.}~\bibnamefont
  {Greschner}}, \ and\ \bibinfo {author} {\bibfnamefont {T.}~\bibnamefont
  {Vekua}},\ }\href {\doibase 10.1103/PhysRevB.85.064420} {\bibfield  {journal}
  {\bibinfo  {journal} {Phys. Rev. B}\ }\textbf {\bibinfo {volume} {85}},\
  \bibinfo {pages} {064420} (\bibinfo {year} {2012})}\BibitemShut {NoStop}%
\end{thebibliography}%
\bibliographystyle{apsrev4-1}

\appendix

\section{Classical $T=0$ phase diagram \label{Classical appendix}}

Our variational ansatz for the classical ground state of \eqref{Hamiltonian} is given by a generalization of \eqref{LT} whereby the cone opening angle $\theta$ depends on $y$.  To see why $\theta_c$ must be the same across all chain spins and $\theta_{IS}$ the same across all interstitial spins, first assume that we know an entire ground-state spin configuration.  Scale $y$ so that the chains lie at integer and the interstitial spins at half-integer $y$, and denote the opening angle of the chain at $y = 0$ by $\theta(0)$.  Now consider the problem of minimizing the Hamiltonian subject to the constraint that the opening angle at $y = 0$ is $\theta(0)$.  This constraint is clearly trivial because it is satisfied by construction, but this formulation of the problem makes it clear by symmetry that $\theta \left( \frac{1}{2} \right) = \theta \left( -\frac{1}{2} \right)$ and $k_y = 0$.  Now considering the problem subject to the constraints that $\theta(0)$, $\theta \left( \frac{1}{2} \right)$, and $\theta \left( -\frac{1}{2} \right)$ take their correct values, we see that $\theta(1) = \theta(-1)$ and so on, showing that $\theta(y)$ must be even in $y$.  But the same reasoning applies if we choose to start at any other chain or row of interstitial spins, so $\theta(y)$ must be symmetric about every integer or half-integer $y$.  This implies that $\theta(n) \equiv \theta_c$ and $\theta(n + \frac{1}{2}) \equiv \theta_{IS}$ for all integer $n$.

We checked numerically that in the region of phase space where the plateau is a local energy minimum of \eqref{LTGSE}, it is also a global minimum.  By extremizing \eqref{LTGSE} with respect to $\theta_c$, $\theta_{IS}$, and $\Delta \phi$, we can get a trigonometric equation in terms of $\Delta \phi$ alone, although the analytic expression expression for the root $\Delta \phi_0$ is extremely complicated.  Off the magnetization plateau, the ground-state cone opening angles are then given by

\begin{align}
\cos \theta_{c,0} &= \frac{h / S}{16 J_2 J'} (- J_1 - 4 J_2 \cos \Delta \phi_0 + J') \csc^4 \left( \frac{\Delta \phi_0}{2} \right) \nonumber \\
\cos \theta_{IS,0} &= \frac{h / S}{8 J_2 J'^2} (J_1 + 4 J_2 \cos \Delta \phi) \\
& \hspace{11pt} \times \left((J_1 - J') \csc^4 \left( \frac{\Delta \phi_0}{2} \right) + 4 J_2 \cot^4 \left( \frac{\Delta \phi_0}{2} \right) \right). \nonumber
\end{align}
The magnetization curve is given by
\beq \label{mofh}
m(h) = S \left( \frac{2}{3} \cos \theta_{c,0} + \frac{1}{3} \cos \theta_{IS,0} \right).
\eeq
We found no evidence of any first-order metamagnetic transitions other than the ferrimagnetic ordering at $h = 0$ in region I of Fig.~\ref{LT phase diagram}.

In regions II and III, the lower critical field $h_l$ is given by the lowest root of the cubic polynomial
\begin{align}
&4 J_2 \left( \frac{h}{S} \right)^3 - \left( (J_1 + 4 J_2)^2 + 24 J_2 J' \right) \left( \frac{h}{S} \right)^2 \nonumber \\
&+ 4 J' \left( 2 (J_1 + 4 J_2)^2 + (-J_1 + 4 J_2) J' ) \right) \frac{h}{S} \nonumber \\
&- 4 J'^2 (2(J_1 + 4 J_2) - J')^2.
\end{align}
In region III, the value of the upper critical field $h_u$ is given by the middle root of the same polynomial.  In regions III and IV, the saturation field $h_\text{sat}$ is given by the highest root of
\begin{align}
&4 J_2 \left( \frac{h}{S} \right)^3 - \left( (J_1 + 4 J_2)^2 + 40 J_2 J' \right) \left( \frac{h}{S} \right)^2 \nonumber \\
& + 4 J' \left( 2 (J_1 + 4 J_2)^2 + (J_1 + 28 J_2) J' \right) \frac{h}{S} \nonumber \\
& - 4 J'^2 \left( 2 (J_1 + 4 J_2) + J' \right).
\end{align}
This value depends very weakly on $J_1$ and $J_2$, and so the saturation field is only slightly greater than $6 S J'$ unless $J_2 \gg |J_1| \gg J'$.

At the boundary $J' = J_1 + 4 J_2$ between regions II and III, the lower critical field assumes the simple expression
\beq
h_l = 2 S J' \left( 1 + \frac{J'}{16J_2} - \sqrt{1 + \left( \frac{J'}{16 J_2} \right)^2} \right)
\eeq
and $h_u = 2 S J'$, so

\begin{align} \label{hlhu}
\frac{h_l}{h_u} &= 1 + \frac{J_1 + 4 J_2}{16J_2} - \sqrt{1 + \left( \frac{J_1 + 4 J_2}{16 J_2} \right)^2} \nonumber \\
& < \frac{5 - \sqrt{17}}{4} = 0.219.
\end{align}

The stabilization curve is given by the second-highest root of the quartic polynomial
\begin{align}
& -54 J_2 J'^4 + 216 J_2 (J_1 + 4 J_2) J'^3 \\
&- (J_1^3 + 276 J_1^2 J_2 + 1776 J_1 J_2^2 + 2240 J_2^3) J'^2 \nonumber \\
&+ 3 (J_1 + 4 J_2)^2 (J_1^2 + 40 J_1 J_2 + 16 J_2^2) J' - 2 J_1 (J_1 + 4 J_2)^4. \nonumber
\end{align}
It approaches $J' = 0.883 J_1 + 2.74 J_2$ for large $J'$ and $J' = (\frac{3}{2} + \sqrt{2})(J_1 + 4 J_2)^2$ for small $J'$.  Along this curve, the critical field at which the plateau vanishes is
\begin{align}
\frac{h_c}{S} = &\left[ 4(J_1 + 4 J_2)^4 J' - 2(J_1 + 4 J_2)^2 (J_1 + 92 J_2) J'^2 \right. \nonumber \\
& \ \left.+ 48 J_2 (5 J_1 + 28 J_2) J'^3 - 72 J_2 J'^4 \right] / \nonumber \\
& \left[(J_1 + 4 J_2)^4 - 48 J_2 (J_1 + 4 J_2)^2 J' \right. \nonumber \\
& \ \left. + 48 J_2 (J_1 + 8 J_2) J'^2 \right]
\end{align}
and the cone-state instability precesses with
\begin{align}
&\cos \Delta \phi_0 = \nonumber \\
&\left[ (J_1 + 4 J_2)^2 (3 J_1 + 4 J_2) \right. \label{stabilizationk0} \\
&\ \left. - 6 \left( J_1^2+20 J_1 J_2 + 32 J_2^2 \right) J' + 3 (J_1 + 32 J_2) J'^2 \right] / \nonumber \\
&\left[ (J_1 - 4 J_2)(J_1 + 4 J_2)^2 - 24 J_2 J' (J_1 - 4 J_2) + 36 J_2 J'^2 \right]. \nonumber
\end{align}
The wave number $\Delta \phi_0$ begins at $0$ at the Lifshitz point and increases monotonically, saturating to $\Delta \phi = 0.989$ rad/site at large $J_2$ and $J'$.

\section{Spin-wave formulation \label{Spin-wave appendix}}

We first rescale the $y$-coordinate to restore isotropy, so that the Fourier coordinate $k_y$ is slightly rescaled relative to the physical crystal momentum and the Brillouin zone becomes isotropic.  We then set $a = 1$ as defined in Fig.~\ref{Hamiltonian diagram}.  The primitive unit cells of the kagom\'{e} lattice form an underlying triangular lattice; we choose the primitive direct lattice vectors $e_1 = (1, \sqrt{3}) / 2$ and $e_2 = (1, -\sqrt{3}) / 2$ and define $e_x := e_1 + e_2$.  The reciprocal lattice vectors $E_1 = 2 \pi (1, 1/\sqrt{3})$ and $E_2 = 2 \pi (1, -1/\sqrt{3})$ satisfy $e_i \cdot E_j = 2 \pi \delta_{ij}$.  We define the components $k_i := \bm{k} \cdot e_i$, so that $\bm{k} = (k_1 E_1 + k_2 E_2)/(2 \pi)$.  We use the Fourier conventions $a_r = \sqrt{\frac{3}{N}} \sum_k e^{i k \cdot r} a(k)$ and similarly (same sign convention) for $b(k)$ and $c(k)$.

The matrix $M(k)$ specifying $H^{(2)}$ is given by
\begin{widetext}
\beq \label{Mk}
M(k) = \left( \begin{array}{ccc}
-2 J_1 - 2 J_2 \left( 1 - \cos k_x \right) + 2 J' + \frac{h}{S} & J_1 \left( 1 + e^{-i k_x} \right) & J' (1 + e^{-i k_1}) \\
J_1 \left( 1 + e^{i k_x} \right) & -2 J_1 - 2 J_2 \left( 1 - \cos k_x \right) + 2 J' + \frac{h}{S} & J' (1 + e^{i k_2}) \\
J' (1 + e^{i k_1}) & J' (1 + e^{-i k_2}) & 4J' - \frac{h}{S}
\end{array} \right).
\eeq
The free-magnon energies are given by the eigenvalues of $SgM(k)$, but with the lowest eigenvalue negated.  Therefore, in order for the plateau to be stable, two eigenvalues of $g M(k)$ must be strictly positive and one must be strictly negative.  To be compatible with the normalization constraint $T g T^\dag = g$, we must order the columns of $T(k)$ such that the eigenvector with the negative-definite eigenvalue goes into the rightmost column, then scale the eigenvectors appropriately.

Along the $k_x$-axis, the highest energy band is given by
\beq \label{aEnergy}
\epsilon_{\tilde{a}}(k_x, 0) = S \left[- 2 J_1 \left( 1 + \cos \left( \frac{k_x}{2} \right) \right) - 2 J_2 (1 - \cos k_x) + 2 J' \right] + h,
\eeq
which is strictly positive for all couplings that can support a classical plateau.  The lower two energy bands are given by \eqref{bcEnergies} with
\begin{align} \label{f1f2}
f_1(k_x) := &\left( -J_1 \left( 1 - \cos \left( \frac{k_x}{2} \right) \right) - J_2 (1 - \cos k_x) - J' \right)^2 \nonumber \\
& + 8 J' \left( -2 J_1 - 4 J_2 \left( 1 + \cos \left( \frac{k_x}{2} \right) \right) + J' \right) \sin^2 \left( \frac{k_x}{4} \right), \\
f_2(k_x) := &J_1 \left( 1 - \cos \left( \frac{k_x}{2} \right) \right) + J_2 (1 - \cos k_x) + J'. \nonumber
\end{align}

The magnons condense at momentum $k = 0$ for sufficiently large $J'$, or else at a root of
\beq \label{k0}
2 J_1^2 + 3 J_1 J_2 + 8 J_2^2 - 3 J_1 J' + J'^2 + 2 J_2 \left( 6 J_1 + 7 J_2 -6 J' \right) \cos \left( \frac{k_x}{2} \right) + J_2 (J_1 + 8 J_2) \cos k_x + 2 J_2^2 \cos \left( \frac{3 k_x}{2} \right).
\eeq

A generic $U(2,1)$ matrix takes the form (supressing an overall phase freedom for each column)
\beq
\left( \begin{array}{ccc}
\cos \theta & \sin \theta \cosh \xi_2 & \sin \theta \sinh \xi_2 \\
-e^{i \delta_1} \sin \theta \cosh \xi_1 & e^{i \delta_1} \left( \cos \theta \cosh \xi_1 \cosh \xi_2 + e^{i \delta_3} \sinh \xi_1 \sinh \xi_2 \right) & e^{i \delta_1} \left( e^{i \delta_3} \sinh \xi_1 \cosh \xi_2 + \cos \theta \cosh \xi_1 \sinh \xi_2 \right) \\
-e^{i \delta_2} \sin \theta \sinh \xi_1 & e^{i \delta_2} \left( \cos \theta \sinh \xi_1 \cosh \xi_2 + e^{i \delta_3} \cosh \xi_1 \sinh \xi_2 \right) & e^{i \delta_2} \left( e^{i \delta_3} \cosh \xi_1 \cosh \xi_2 + \cos \theta \sinh \xi_1 \sinh \xi_2 \right)
\end{array} \right).
\eeq

For the Bogoliubov transformations $T(k)$ for this system, $\delta_3 = 0$ or $\pi$.  For $k_y = 0$, \eqref{atilde} gives that $\xi_1 = 0$, $\theta = \pi / 4$, and $\delta_1 = k_x / 2$.  Since $\xi_1 = 0$ we can without loss of generality set $\delta_3 = 0$, and diagonalizing $gM(k)$ gives $\delta_2 = k_x / 4$ and equations \eqref{T} and \eqref{xi}.  \eqref{xi} can be equivalently rewritten as
\beq \label{xi2}
\tanh \xi(k_x) = \frac{\epsilon_{\tilde{c}}(k_x, 0)|_{h=0} - 4 S J'}{2 \sqrt{2} S J' \cos \left( \frac{k_x}{4} \right)}.
\eeq

At full saturation, the matrix $M_\text{sat}$ is given by \eqref{Mk} with all the $J'$ in the main diagonal and the $h/S$ in the $(3,3)$ element negated.  The energy band $\epsilon_{\tilde{a}, \text{sat}}(k_x, 0) = \epsilon_a(k_x, 0) - 4 S J'$ where $\epsilon_{\tilde{a}}$ is given in \eqref{aEnergy}; the $\tilde{a}$ magnons for the fully polarized state and for the plateau state correspond to identical states on the chains, with only the interstitial spins flipped.  The other two bands are $\epsilon_{\tilde{b}/\tilde{c}}(k_x, 0) = S \left( \pm \sqrt{f_{1, \text{sat}}(k_x)} - f_{2, \text{sat}}(k_x) \right) + h$ with $f_{2, \text{sat}}(k_x) = f_2(k_x) + 2 J'$ and 
\begin{align}
f_{1, \text{sat}}(k_x) = &\left(-J_1 \left(1 - \cos \left(\frac{k_x}{2} \right) \right) - J_2 (1 - \cos k_x) - 3 J' \right)^2 \\
&+ 8 J' \left(-2 J_1 - 4 J_2 \left( \cos \left( 1 + \frac{k_x}{2} \right) \right) - J' \right) \sin^2 \left( \frac{k_x}{4} \right). \nonumber
\end{align}
The saturation field is given by the value of $h$ at which the lowest of these three bands touches zero; we can show that this is always the $\epsilon_{\tilde{c}}$ band.  For $J' \geq J_1 + 4 J_2,\ \epsilon_{\tilde{c}}(k_x, 0)$ is minimized at $k_\text{sat} = 0$ and $h_\text{sat} = 6 S J'$, and for $J' < J_1 + 4 J_2$ it is minimized at a root of
\beq
2 J_1^2 + 3 J_1 J_2 - J_1 J' + 8 J_2^2 - J'^2 + J_2 \left( 2 (6 J_1 + 7 J_2 - 2 J') \cos \left( \frac{k_x}{2} \right) + (J_1 + 8 J_2) \cos k_x + 2 J_2 \cos \left( \frac{3 k_x}{2} \right) \right).
\eeq

\end{widetext}

\section{Magnon self-energy corrections to plateau critical fields \label{Corrections app}}

The Fourier- and Bogoliubov-transformed quartic Hamiltonian $H^{(4)}$ consists of a sum over four momenta, one of which is fixed by momentum conservation.  Normal-ordering produces another delta-function constraint, reducing the number of independent momenta to two: the usual crystal momentum $k$ corresponding to lattice translational invariance, whose value is fixed to the momentum of the free-magnon instability at the critical field, and a dummy momentum $k'$ which is summed over.  Thus the matrix elements of $\Delta M(k)$ will be given by very complicated Fourier transforms of the matrix elements of the Bogoliubov transformation $T(k)$.  $k'$ is summed over the entire 2D Brillouin zone, not just the $k_x$-axis, so we cannot use $\eqref{T}$ but must calculate $T(k)$ numerically.  $\Delta M_{12}$, $\Delta M_{13}$, $\Delta M_{21}$, and $\Delta M_{31}$ are all zero, and $\Delta M_{23} = \Delta M_{32}$.

$T(k)$ has certain nonobvious symmetries that result in the following identities relating various Fourier transforms (where we denote $\IFT[f(k')](x) := \frac{1}{N} \sum_{k' \in \text{BZ}} f(k')\, e^{i k' \cdot x}$, $i = 1, 2, 3$, and $a = 1, 2$):
\begin{itemize}
\item $\IFT[T_{3a}^*(k')\, T_{1a}(k')](0) = \IFT[T_{3a}^*(k')$ $T_{2a}(k')](0) $ \\
$= \IFT[T_{3a}^*(k')\, T_{1a}(k')](e_1) $ \\
$= \IFT[T_{2a}^*(k')\, T_{3a}(k')](e_2)$
\item $\IFT[T_{13}^*(k')\, T_{33}(k')](0) = \IFT[T_{23}^*(k')$ $T_{33}(k')](0) $ \\
$= \IFT[T_{33}^*(k')\, T_{13}(k')](e_1) $ \\
$= \IFT[T_{23}^*(k')\, T_{33}(k')](e_2)$
\item $\IFT[T_{2a}^*(k')\, T_{1a}(k')](0) = \IFT[T_{2a}^*(k')\, T_{1a}(k')](e_x)$
\item $\IFT[T_{13}^*(k')\, T_{23}(k')](0) = \IFT[T_{23}^*(k')\, T_{13}(k')](e_x)$
\item $\IFT[|T_{1i}(k')|^2](e_x) = \IFT[|T_{2i}(k')|^2](e_x)$
\item $\IFT[|T_{13}(k')|^2](0) = \IFT[|T_{23}(k')|^2](0)$
\end{itemize}
These identites allow us to simplify the matrix elements of $M(k)$ to
\begin{widetext}
\begin{align}
\Delta \epsilon_{\tilde{b}}(k_x, 0) &= \Delta M_{22} = \frac{1}{N} \sum_{k' \in BZ} \left[ \cosh^2 \xi\, \left( f_1(k')\, e^{i k'_x} + f_2(k') \right) + J' \left( \sinh(2 \xi)\, f_3(k') + \sinh^2 \xi\, f_4(k') + f_5(k') \right) \right] \label{delta es} \\
\Delta \epsilon_{\tilde{c}}(k_x, 0) &= \Delta M_{33} = \frac{1}{N} \sum_{k' \in BZ} \left[ \sinh^2 \xi \left( f_1(k')\, e^{i k'_x} + f_2(k') \right) + J' \left( \sinh(2 \xi)\, f_3(k') + \cosh^2 \xi\, f_4(k') - f_5(k') \right) \right] \nonumber
\end{align}
where $\xi(k_x)$ is given by \eqref{xi}, \eqref{bcEnergies}, and \eqref{f1f2}, and
\begin{align}
f_1(k_x, T(k')) :=&\, J_2 \left( 1 - e^{-i k_x} \right) \left( -|T_{13}|^2 + e^{i k_x} \left( | T_{11}|^2 + |T_{12}|^2 \right) \right) \nonumber \\
f_2(k_x, T(k')) :=&\, -J_1 \left( 1 - e^{-\frac{1}{2} i k_x} \right) \left(e^{\frac{1}{2} i k_x} \left( |T_{13}|^2 - T_{11} T_{21}^* - T_{12} T_{22}^* \right) - |T_{13}|^2 + T_{23} T_{13}^* \right) \nonumber \\
& + 2 J_2 (1 - \cos(k_x)) |T_{13}|^2 - J' \left(2 |T_{31}|^2 + 2 |T_{32}|^2 + 3 \left( T_{11} T_{31}^*+ T_{12} T_{32}^* + T_{33} T_{13}^* \right) \right) \nonumber \\
f_3(k_x, T(k')) :=&\, \frac{1}{\sqrt{2}} \left( -\cos \left( \frac{k_x}{4} \right) \left(2 |T_{13}|^2 + 2 |T_{31}|^2 + 2 |T_{32}|^2 + T_{11} T_{31}^* + T_{12} T_{32}^* + T_{33} T_{13}^* \right) \right. \\
& \left. \hspace{30pt} -e^{\frac{1}{4} i k_x} \left(T_{11} T_{31}^* + T_{12} T_{32}^* \right) - e^{-\frac{1}{4} (i k_x)} T_{33} T_{13}^* \right) \nonumber \\
f_4(T(k')) :=&\, -4 |T_{13}|^2 \nonumber \\
f_5(T(k')) :=&\, 2 \left( T_{11} T_{31}^* + T_{12} T_{32}^* + T_{33} T_{13}^* \right). \nonumber
\end{align}
\end{widetext}

For large $J'$, the leading plateau instabilities occur at $k = 0$ and the critical fields take on very simple forms: for $J' \geq 2(J_1 + 4 J_2),\ h_l \equiv 0$ and for $J' \geq J_1 + 4 J_2,\ h_u = 2 S J'$.  In these regions the quantum corrections become particularly simple, because the Bogoliubov transformation becomes independent of the Hamiltonian parameters.  $\xi(k = 0) = -\arccoth(\sqrt{2}) \approx -0.881$ and
\beq
T(0) = \left( \begin{array}{ccc}
\frac{1}{\sqrt{2}} & 1 & -\frac{1}{\sqrt{2}} \\
-\frac{1}{\sqrt{2}} & 1 & -\frac{1}{\sqrt{2}} \\
0 & -1 & \sqrt{2} \end{array} \right).
\eeq
$\Delta \epsilon_{\tilde{b}}(0)$ vanishes identically, so for $J' \geq 2(J_1 + 4 J_2)$, the $o(S^0)$ quantum corrections do not shift the lower critical field -- this is an exact result that holds to all orders, as discussed in the text.  $\Delta \epsilon_{\tilde{c}}(0)$ reduces to
\begin{align}
\Delta \epsilon_{\tilde{c}}(0) = -J' \times \frac{1}{N} \sum_{k' \in BZ} &\left[ 4 |T_{13}|^2 - 2 |T_{32}|^2 - 2 |T_{32}|^2 \right. \nonumber \\
&\left. + T_{11} T_{32}^* + T_{12} T_{32}^* + T_{33} T_{13}^* \vphantom{|T_{13}|^2} \right],
\end{align}
where $T$ is shorthand for $T(k')$.  This expression has a weak implicit dependence on $J_1$ and $J_2$ through $T(k')$, but it depends much more strongly on $J'$.

To find the behavior of $h_l$ as $J' \rightarrow 2(J_1 + 4 J_2)^-$, we can expand \eqref{k0} to second order in $k_x$ and solve to get
\begin{align}
k_c &\approx \pm \sqrt{\frac{((J_1 + 4 J_2) - J') (2 (J_1 + 4 J_2) - J')}{J_2 \left(2 (J_1 + 4 J_2) - \frac{3}{2} J' \right)}} \nonumber \\
k_l &\approx \pm \sqrt{\frac{2(J_1 + 4 J_2) - J'}{J_2}}.
\end{align}
Then \eqref{bcEnergies} and \eqref{f1f2} give that $h_l$ vanishes quadratically as $J' \rightarrow 2(J_1 + 4 J_2)^-$:
\beq
h_l = -\epsilon_{\tilde{b}}(k_l, 0)|_{h = 0} \approx \frac{S}{4 J_2} ( 2(J_1 + 4 J_2) - J')^2.
\eeq

For fixed $J_2$, the maximum value of $S \Delta h_l / h_l$ in Fig.~\ref{CorrectionsPlot} (a) appear to occur along the line $J' = J_1 + 4 J_2$.  In the limit $J_2, J' \gg |J_1|$, this value approaches $-0.73$. 

$\xi(k_u) \equiv -\arccoth(\sqrt{2})$ for $J' \geq J_1 + 4 J_2$, while $\xi(k_l) \equiv -\arccoth(\sqrt{2})$ for $J' \geq 2(J_1 + 4 J_2)$.  $\xi_l$ and $\xi_u$ both diverge logarithmically to $-\infty$ along the stabilization curve.  For fixed $J'$,
\beq
e^{-\xi(k_c)} \sim \left( \frac{1}{\sqrt{2}} \sin \left( \frac{k_c}{2} \right) \sin \left( \frac{k_c}{4} \right) \frac{J_{2c} - J_2}{J'} \right)^{-\frac{1}{4}}
\eeq
and the coherence factors $\cosh \xi$ and $\sinh \xi$ diverge as $\pm \frac{1}{2} e^{-\xi(k_c)}$.

In the case of full saturation, we have that for $J' \geq J_1 + 4 J_2$ the leading instabilities occur at $k = 0$, which corresponds to $\theta(0) = \arctan \left( 1 / \sqrt{2} \right) = 0.615$ and
\beq
T(0) = \left( \begin{array}{ccc}
\frac{1}{\sqrt{2}} & \frac{1}{\sqrt{3}} & \frac{1}{\sqrt{6}} \\
-\frac{1}{\sqrt{2}} & \frac{1}{\sqrt{3}} & \frac{1}{\sqrt{6}} \\
0 & \frac{1}{\sqrt{3}} & -\sqrt{\frac{2}{3}} \end{array} \right).
\eeq
$\theta(k_\text{sat}) \in \left[ \arctan \left( 1 / \sqrt{2} \right), \pi / 2 \right)$ remains finite over the entire phase diagram.

\section{Magnon interactions \label{Interactions app}}

Away from the stabilization curve, the repulsion between antiparallel magnons is always stronger than the repulsion between parallel magnons.

At full saturation, the $\tilde{c}$ magnons repel with energy $4 J'$ for $J' > J_1 + 4 J_2$; for $J' < J_1 + 4 J_2$ the repulsion depends slightly on $J_1$ and $J_2$ but is still approximately $4J'$ for both the parallel and antiparallel scattering processes.  The parallel process's energy dominates except at very small $J'$.  Therefore, quantum effets do not affect the saturation field $h_\text{sat}$ at this order in $1/S$.

\begin{widetext}
The matrix $\hat{M}(k)$ is defined as
\beq \label{Mhat}
\hat{M}(k) = \left( \begin{array}{ccc}
-2 J_1 - 2 J_2 \left( 1 - \cos k_x  \right) + 2 J' + \frac{h}{S} & 2 J_1 \cos \left( \frac{k_x}{2} \right) & 2 J' \cos \left( \frac{k_1}{2} \right) \\
2 J_1 \cos \left( \frac{k_x}{2} \right) & -2 J_1 - 2 J_2 \left( 1 - \cos k_x \right) + 2 J' + \frac{h}{S} & 2 J' \cos \left( \frac{k_2}{2} \right) \\
2 J' \cos \left( \frac{k_1}{2} \right) & 2 J' \cos \left( \frac{k_2}{2} \right) & 4J' - \frac{h}{S}
\end{array} \right),
\eeq
\end{widetext}
which we abbreviate to
\beq
\left( \begin{array}{ccc}
\alpha & \beta & \gamma_+ \\
\beta & \alpha & \gamma_- \\
\gamma_+ & \gamma_- & \delta
\end{array} \right).
\eeq

Only $\gamma_\pm = 2 J' \cos \left( \frac{k_x \pm \sqrt{3} k_y}{4} \right)$ depends on $k_y$; if $k_y = 0$ then $\gamma_+ = \gamma_- = \gamma := 2 J' \cos(k_x / 4)$ and the matrix $A(k)$ in \eqref{H2eff} is given by
\beq \label{A}
A(k_x, 0) := \left( \begin{array}{cc} \alpha + \beta & \sqrt{2} \gamma \\ \sqrt{2} \gamma & \delta \end{array} \right).
\eeq

$\xi(k_c)$ diverges to $-\infty$ at the critical momenta, and $h_c = \epsilon_{\tilde{c}}(\pm k_c, 0)|_{h = 0}$, so \eqref{xi} and \eqref{xi2} give
\begin{align}
2 \sqrt{2} J' \cos( k_x/4) &= 4 J' - h_c /S \\
&= -J_1 \left(1 - \cos \left( \frac{k_x}{2} \right) \right) \nonumber \\
&\hspace{11pt} - J_2 (1 - \cos k_x) + 3 J' \nonumber \\
\sqrt{2} \gamma &= \delta = \frac{1}{2} (\alpha + \beta + \delta) = \alpha + \beta \nonumber
\end{align}
so $A_\alpha(0) = \delta \left( \begin{array}{cc} 1 & 1 \\ 1 & 1 \end{array} \right)$.

In step 4 of the procedure given in Table~\ref{Coarse-graining}, $\Phi(k, \tau) = \left( \begin{array}{c} \phi(k, \tau) \\ \eta(k, \tau) \end{array} \right) = U^\dag(k) \Omega(k, \tau)$ diagonalizes $A(k)$ to
\beq
D(k) := S U^\dag A U = \left( \begin{array}{cc} \epsilon_\phi(k) & 0 \\ 0 & \epsilon_\eta(k) \end{array} \right),
\eeq
with $\epsilon_{\phi,\alpha}(0) = 0$, $\epsilon_{\eta,\alpha}(0) = 2 S \delta$, and $U_\alpha(0)$ given by \eqref{U}.  Action \eqref{SfreeOmega} becomes
\beq
S = \int \frac{d^2k}{(2 \pi)^2} d\tau \left( \Phi^\dag U^\dag g U \partial_\tau \Phi + \Phi^\dag D \Phi \right).
\eeq

In step 5, the imaginary time derivative term expands to
\beq
U_\alpha(k)^\dag g U_\alpha(k) = \left( \begin{array}{cc} 0 & 1 \\ 1 & 0 \end{array} \right) \pm \left( \begin{array}{cc} 1 & 0 \\ 0 & -1 \end{array} \right) \lambda k_x + o(k^2),
\eeq
where the top line corresponds to $R$ and the bottom to $L$, and $\lambda := \left( J_1 \sin(k_c / 2) + 2 J_2 \sin k_c \right) / \Delta_\eta$ ranges from $0$ to $0.135$.  The second term in the action expands to
\beq
\sum_{\alpha = R, L} S \bigg( \Delta_\eta \eta^\dag_\alpha \eta_\alpha + \phi^\dag_\alpha \phi_\alpha \sum_{i = x, y} \kappa_i k_i^2 \bigg),
\eeq
where
\beq
\kappa_i := \frac{1}{2S} \left( \frac{\partial^2 \epsilon_\phi}{\partial k_i^2} \right) \bigg|_{k = (\pm k_c, 0)}.
\eeq
$\epsilon_\phi(k_x, 0) = \frac{S}{2} \left( \alpha + \beta + \delta - \sqrt{8 \gamma^2 - (\alpha + \beta - \delta)^2} \right)$, so $\kappa_x$ is straightforward to calculate.  $\kappa_y$ is not, because we do not have a closed-form expression for the gapless mode's energy off the $k_x$-axis.  Exanding $\gamma_\pm$ to second order in $k_y$ about the critical momenta gives
\beq
\gamma_\pm = \gamma \mp \frac{1}{4} \sqrt{3(4 J'^2 - \gamma^2)}\, k_y - \frac{3}{32} \gamma\, k_y^2 + o(k_y^3).
\eeq
Since $a_-$ is no longer an eigenstate if $k_y \neq 0$, we must directly diagonalize $\hat{M}(k)$:
\begin{align}
\det(S \hat{M} - \epsilon) &= \frac{3}{8} \delta \left( \alpha \delta - 4 J'^2 \right) k_y^2 \\
&\hspace{11pt} + \left( -4 \alpha \delta +\frac{13}{8} \delta^2 + \frac{3}{2} J'^2 k_y^2 \right) \epsilon + o(\epsilon^2). \nonumber
\end{align}
We only need the leading-order shift in the zero root, so we can discard the $o(\epsilon^2)$ terms.  The matrix becomes singular when
\beq
\epsilon_\phi = \frac{3S \left(\alpha  \delta -4 J'^2 \right)}{32 \alpha -13 \delta} k_y^2 + o(k_y^3)
\eeq
so $\kappa_y = 3 \left(\alpha \delta - 4 J'^2 \right) / \left(32 \alpha - 13 \delta \right)$.  The $\kappa_i$ (and therefore $c_i$ as well) are close in value and grow approximately proportionately to $J'$ along the stabilization curve.

Steps 1 and 2 of Table~\ref{Coarse-graining} combine to
\beq
\Psi(k_x) = \left( \begin{array}{ccc}
\frac{1}{\sqrt{2}} & \frac{1}{\sqrt{2}} & 0 \\ -\frac{e^{\frac{i k_x}{2}}}{\sqrt{2}} & \frac{e^{\frac{i k_x}{2}}}{\sqrt{2}} & 0 \\ 0 & 0 & e^{\frac{i k_x}{4}}
\end{array} \right) \left( \begin{array}{c} a_-(k_x) \\ d(k_x) \\ \bar{c}^\dag(-k_x) \end{array} \right).
\eeq

$V_p$ and $V_a$ in \eqref{SPhi} are given by
\begin{align}
V_p &:= (S \Delta_\eta a)^2 \frac{3}{4} \left[ J_1 \left( 1 - \cos \left( \frac{k_c}{2} \right) \right) + J_2 (1 - \cos k_c) \right. \nonumber \\
& \hspace{65pt} \left. + J' \left(-4 + 3 \sqrt{2} \cos \left( \frac{k_c}{4} \right) \right) \right] \label{Vs} \\
V_a &:=\ (S \Delta_\eta a)^2 \left[ \frac{3}{4} J_1 \left( -2 \cos \left( \frac{k_c}{2} \right) + \cos k_c + 1 \right) \right. \nonumber \\
&\hspace{62pt}- 3 J_2 \sin^2 \left( \frac{k_c}{2} \right) \cos k_c \nonumber \\
& \hspace{62pt} \left. + \frac{3}{2} J' \cos \left( \frac{k_c}{4} \right) \left(3 \sqrt{2} - 4 \cos \left( \frac{k_c}{4} \right) \right) \right]. \nonumber
\end{align}

Operators linear in $\Psi$ are trivially normal-ordered, so we can combine steps 1-5 in Table~\ref{Coarse-graining} to
\beq
\langle \Psi_R \rangle = \frac{1}{2} \left( \begin{array}{ccc} \sqrt{2} & 1 & 1 \\ -\sqrt{2} e^{\frac{1}{2} i k_c} & e^{\frac{1}{2} i k_c} & e^{\frac{1}{2} i k_c} \\ 0 & -\sqrt{2} e^{\frac{1}{4} i k_c} & \sqrt{2} e^{\frac{1}{4} i k_c} \end{array} \right) \left( \begin{array}{c} \langle a_{-R} \rangle \\ \langle \phi_R \rangle \\ \langle \eta_R \rangle \end{array} \right)
\eeq
and similarly for $\Psi_L$ with $k_c \to -k_c$, so
\begin{align}
\langle a(k_c, 0) \rangle &= \frac{1}{2} \langle \phi_R \rangle = \frac{1}{2} a^2 \sqrt{S \Delta_\eta \rho_R} e^{i \theta_R}, \\
\langle b(k_c, 0) \rangle &= \frac{1}{2} e^{\frac{1}{2} i k_c} \langle \phi_R \rangle = \frac{1}{2} a^2 \sqrt{S \Delta_\eta \rho_R} e^{i \left( \frac{1}{2} k_c + \theta_R \right)}, \nonumber \\
\left \langle c(k_c, 0) \right \rangle &= -\frac{1}{\sqrt{2}} e^{\frac{1}{4} i k_c} \left \langle \phi_L^\dag \right \rangle = -a^2 \sqrt{\frac{S \Delta_\eta \rho_L}{2}} e^{i \left( \frac{1}{4} k_c - \theta_L \right)}. \nonumber
\end{align}
and the corresponding expressions with $k_c \to -k_c$ and $R \leftrightarrow L$.  (The contribution to the phase shift in $\langle b \rangle$ and $\langle c \rangle$ proportional to $k_c$ comes from the real-space sublattice displacements.)  Combining these equations with $\langle S_b^+(r) \rangle = \big( \sqrt{2 S} / a \big) e^{i k_c x} \langle b(k_c, 0) \rangle$ and similar expressions for the $a$ and $c$ sublattices gives \eqref{Splus}.

\section{RG analysis of action near stabilization curve \label{RG app}}

The $\chi$ field only contributes significantly to the path integral if the potential energy term $\int d^3x\, \frac{1}{2} m^2 \chi^2 \lesssim 1$, so $\chi$ effectively only varies within a width $\Delta \chi \sim 1/m \sim \sqrt{V_a - V_p}$.  Near the $O(4)$ point $V_a = V_p$, $\chi$ is small so
\begin{align}
e^{-\Delta S} \approx \ &1 + \int d^3x \left[ g \chi(x) \langle O(x) \rangle_{S_{O(4)}} \right] \\
&+ \int d^3x\, d^3 x' \left[ g^2 \chi(x) \chi(x') \langle O(x) O(x') \rangle_{S_{O(4)}} \right]. \nonumber
\end{align}
The second term of the RHS vanishes by the $O(4)$ symmetry, and expanding the negative logarithm gives \eqref{DeltaS}.

Ref.~\onlinecite{Calabrese03} refers to $\frac{1}{2} O$ as $\mathcal{P}_{2,2}$ because it transforms as the $l = 2,\ m = 2$ representation of $O(4)$.  If we add a symmetry-breaking perturbation $h_p \mathcal{P}_{2,2}$ to the $O(4)$ action, then in the notation of Ref.~\onlinecite{Calabrese03}, the singular part of the free energy density at small $h_p$ and reduced temperature $t$ goes as
\beq
\mathcal{F}_\text{sing}(t, h_p) \approx |t|^{d \nu} \hat{\mathcal{F}}(h_p |t|^{-y_{2,2} \nu}),
\eeq
where $y_{2,2}$ is the RG dimension of $\mathcal{P}_{2,2}$ and $\nu = 0.749$ is the $O(4)$ critical exponent.  The standard scaling relations $\Delta = \frac{1}{2} (d - 2 + \eta)$ and $\eta = d + 2 - 2y_{2,2}$ give $\Delta = d - y_{2,2}$.  Ref.~\onlinecite{Calabrese03}'s five-loop $\epsilon$-expansion gives $y_{2,2} = 1.813(6)$, so $\Delta = 1.186(4)$.

If we normalize $f$ appropriately, then the Fourier transform in \eqref{DeltaSFT} can be expanded as
\beq
\int d^3y \left( e^{-i \xi k y} \frac{f(|y|)}{|y|^{2 \Delta}} \right) = 1 - \frac{1}{2} b\, (\xi k)^2 + o \left( (\xi k)^4 \right),
\eeq
and setting $g^2 = \xi^{2 \Delta - 5}/(b C)$ gives \eqref{DeltaSNormalized}.  The coefficient
\begin{align}
b &= -\nabla^2 \big|_{\xi k = 0} \IFT \left[ \frac{f(|y|)}{|y|^{2 \Delta}} \right](\xi k) \label{b} \\
&= \IFT \left[ |y|^2 \frac{f(|y|)}{|y|^{2 \Delta}} \right](0) = \int d^3y \left( \frac{f(|y|)}{|y|^{2 (\Delta - 1)}} \right), \nonumber
\end{align}
is presumably positive, so the kinetic term is bounded below.

\eqref{Stilde} can be written as
\begin{align}
\tilde{S} &= \frac{1}{2} \int d^3x \left[ \vec{\varphi}_R \cdot P(1 - P^{-1} K) \vec{\varphi}_R \right. \\
&\hspace{53pt} \left. + \vec{\varphi}_L \cdot P(1 + P^{-1} K) \vec{\varphi}_L \right]. \nonumber
\end{align}
We can integrate over $\vec{\varphi}_\alpha$ and absorb $\det P$ into the proportionality constant because it does not depend on $\chi$:
\beq
e^{-\Delta S} \propto \det \left( 1 - P^{-1} K \right)^{-1/2} \det \left( 1 + P^{-1} K \right)^{-1/2}.
\eeq
$\ln \left( \det A \right) = \Tr \left( \ln A \right)$ so
\beq
\Delta S = \frac{1}{2} \Tr \left[ \ln \left( 1 - \left( P^{-1} K \right)^2 \right) \right] + \text{const.}
\eeq
and expanding the logarithm gives \eqref{DeltaSTrace}.

Series \eqref{Uchi} can be summed to the closed-form expression
\begin{align}
U(\chi) =& -\frac{r^{3/2}}{12 \pi} \left[ \sqrt{1 + \frac{2 g \chi}{r}} + \sqrt{1 - \frac{2 g \chi}{r}} - 2 \right. \\
&\hspace{38pt} \left. + \frac{2 g \chi}{r} \left( \sqrt{1 + \frac{2 g \chi}{r}} - \sqrt{1 - \frac{2 g \chi}{r}} \right) \right], \nonumber
\end{align}
if $|g \chi/r| \leq 1/2$, but this expression is only valid for $g \chi / r \ll 1$ because the neglected $\vec{\varphi}^4$ coupling modifies $U(\chi)$ for large $\chi$, as explained in the main text.

\section{Chiral liquid wavefunction \label{Wavefunction app}}

In terms of the condensed magnon operators $\Psi_\alpha$,
\beq
\langle J^z_{s s'}(\Delta r) \rangle = \frac{S}{a^2} \sum_\alpha \big \langle \Psi^\dag_\alpha B^{(s s')}_\alpha(\Delta r) \Psi_\alpha \big \rangle,
\eeq
where $B^{(s s')}_\alpha(\Delta r)$ is a $3 \times 3$ matrix.  The only nonzero component of $B^{(ss)}_R(\Delta r)$ is $\left( B_R^{(ss)} \right)_{ss} = 2 \sin(k_c \Delta x)$.  For $s < s'$, the nonzero components of $B_R^{(ss')}(\Delta r)$ are $\left( B_R^{(ss')} \right)_{ss'} = \exp[i(k_c \Delta x - \pi/2)]$ and $\left( B_R^{(ss')} \right)_{s's} = \left( B_R^{(ss')} \right)_{ss'}^*$.  $B_L^{(ss')}(\Delta r)$ is given by $B_R^{(ss')}(\Delta r)$ with $k_c \to -k_c$.  Note that $B_\alpha^{(ss')}(\Delta r) = B_\alpha^{(s's)}(-\Delta r)$ because $J^z_{ij}$ is antisymmetric.  The procedure in Table~\ref{Coarse-graining} gives \eqref{Jz} if we factor in the physical displacement of the sublattices.

\end{document}